\documentclass[]{jfm}

\usepackage{amsmath, mathtools, commath, mathrsfs, esint}
\usepackage{natbib}
\usepackage{graphics, tikz, overpic, rotating, indentfirst, multirow}
    \graphicspath{{./figures/}}
    \usetikzlibrary{positioning, petri, shapes, arrows, calc}
    \tikzset{>=stealth}

\newcommand{\RomanNumeralCaps}[1]
\linenumbers

\renewcommand{\vec}[1]{\boldsymbol{#1}}

\newcommand{\bn}{\vec{\nabla}}
\newcommand{\ee}{\mathrm{e}}
\newcommand{\ii}{\mathrm{i}}

\newcommand{\EQ}{\begin{equation}}
\newcommand{\EN}{\end{equation}}

\newcommand{\T}{^{\mathrm{T}}}
\renewcommand{\norm}[1]{\| #1 \|}
\newcommand{\widebar}[1]{\mskip.5\thinmuskip\overline{\mskip-.5\thinmuskip {#1} \mskip-.5\thinmuskip}\mskip.5\thinmuskip}  

\newcommand{\red}[1]{\textcolor{red}{#1}}

\newcommand{\white}[1]{\textcolor{white}{#1}}

\definecolor{dgreen}{rgb}{0.0,0.55,0.0}
\definecolor{dMagenta}{rgb}{1,0,1}

\shorttitle{Dynamics of the spin Euler equation}
\shortauthor{Z. Meng and Y. Yang}

\title{Lagrangian dynamics and regularity of the spin Euler equation}

\author{Zhaoyuan Meng\aff{1}
    \and
    Yue Yang\aff{1,2}\corresp{\email{yyg@pku.edu.cn}}}

\affiliation{\aff{1}State Key Laboratory for Turbulence and Complex Systems, College of Engineering, Peking University, Beijing 100871, PR China
\aff{2}HEDPS-CAPT, Peking University, Beijing 100871, PR China}

\begin{document}
\maketitle

\begin{abstract}
We derive the spin Euler equation for ideal flows by applying the spherical Clebsch mapping.
This  equation is based on the spin vector rather than the velocity.
It enables a feasible Lagrangian study of fluid dynamics, as the isosurface of a spin-vector component is a vortex surface and material surface in ideal flows.
The spin Euler equation is also equivalent to a special case of the Landau--Lifshitz equation with a specific effective magnetic field, revealing a possible connection between ideal flow and magnetic crystal.
We conduct direct numerical simulations of three ideal flows of the vortex knot, vortex link and modified Taylor--Green flow by solving the spin Euler equation.
The evolution of the Lagrangian vortex surface illustrates that the regions with large vorticity are rapidly stretched into spiral sheets.
We establish a non-blowup criterion for the spin Euler equation, suggesting that the Laplacian of the spin vector must diverge if the solution forms a singularity at some finite time.
%
The DNS result exhibits a pronounced double-exponential growth of the maximum norm of Laplacian of the spin vector, showing no evidence of the finite-time singularity formation if the double-exponential growth holds at later times.
Moreover, the present criterion with Lagrangian nature appears to be more sensitive than the Beale--Kato--Majda criterion in detecting the flows that are incapable of producing finite-time singularities.
\end{abstract}


\section{Introduction}\label{sec:intro}
The dynamics of an ideal flow is governed by the three-dimensional (3D) incompressible Euler equation.
One of the outstanding open problems in fluid mechanics is whether smooth initial data can lead to finite-time singularities in the ideal flow.
This problem is closely related to the existence and smoothness of solutions to the Navier--Stokes (NS) equation~\citep[e.g.][]{Fefferman2001_Existence,Doering2009_The, Wei2016_Regularity, Ayala2017_Extreme}.

The regularity of the incompressible Euler equation has been extensively studied.
Various criteria for blowup and non-blowup, based on different quantities and techniques, have been reviewed by \citet{Chae2008_Incompressible}, \citet{Gibbon2008_The} and \citet{Drivas2022_Singularity}.
Several criteria relate the occurrence of singularity to the growth of the vorticity $\vec{\omega}$ which plays a vital role in fluid dynamics.
The Beale--Kato--Majda (BKM) criterion establishes a sufficient condition for the regularity in terms of $\vec{\omega}$ \citep{Beale1984_Remarks}.
The geometric criterion of \citet{Constantin1996_Geometric} relates the regularity of the velocity to the smoothness of the vorticity direction.
Moreover, there are some refined analytical criteria for blowup \citep[e.g.][]{Planchon2003_An, Zhou2013_Logarithmically}.

The regularity of the incompressible Euler equations has also been investigated by large-scale numerical simulations.
\citet{Brachet1983_Small, Brachet1992_Numerical} and \citet{Bustamante2012_Interplay} conducted numerical studies of the evolution of the inviscid Taylor--Green (TG) flow,
and showed a near-exponential growth of the maximum vorticity over time, with regions of the high vorticity predominantly confined within thin, sheet-like structures.
The formation of vortex sheets reduces the three-dimensionality, which suppresses the formation of a finite-time singularity~\citep{Constantin1996_Geometric, Drivas2022_Singularity}.

As the regularity of the 2D Euler equations was established \citep{Pumir1990_Collapsing, Majda2002_Vorticity, Ohkitani2008_A}, subsequent numerical studies primarily focused on the carefully designed initial condition that would enhance the vorticity growth.
%
However, different vorticity growth trends were observed in the numerical simulations with different initial conditions or even the same initial condition.

For the two perturbed anti-parallel vortex tubes, \citet{Kerr1993_Evidence, Kerr2005_Velocity} found $\norm{\vec{\omega}}_\infty\sim(t_0-t)^{-1}$ which provided a strong evidence in favour of blowup, whereas \citet{Hou2007_Computing} and \citet{Hou2009_Blow} obtained a high-resolution numerical solution that is still regular beyond the presumed blowup time $t_0$ and exhibited a maximum vorticity growth slower than double-exponential.
The analysis was subsequently revisited in \citet{Bustamante2008_3D} which proposed a hypothesis of a vorticity growth of $\norm{\vec{\omega}}_\infty\sim (t_0-t)^{-\gamma}$ with $\gamma>1$, and in \citet{Kerr2013_Bounds} which reported a double-exponential growth.

The vorticity growth of $\norm{\vec{\omega}}_\infty\sim(t_0-t)^{-1}$ was also observed in \citet{Grauer1998_Adaptive} using a perturbed cylindrical shear flow, and in \citet{Orlandi2012_Vortex} using the collision of two Lamb dipoles.
\citet{Agafontsev2015_Development, Agafontsev2017_Asymptotic} reported that the vorticity grows exponentially in time in a shear flow with random perturbations.
Moreover, \citet{Ricca1999_Evolution} suggested that the vortex knot is a useful configuration for studying singularity formation, and also pointed out the lack of study on the evolution of vortex knots or links with finite thickness in ideal flows.



Several studies examined the Kida--Pelz (KP) flow \citep{Kida1985_Three, Boratav1994_Direct, Pelz2001_Symmetry}, which is another highly symmetric flow for investigating the formation of potential finite-time singularity.
\citet{Grafke2008_Numerical} compared different numerical methods applied to a KP flow in spectral and real spaces,
and found no evidence of blowup at the times predicted by previous studies, which was confirmed by \citet{Hou2008_Blowup}.
They also observed that the vorticity increases exponentially along the Lagrangian trajectory. 
%

Furthermore, there are several studies which are not based on the Euler equation for investigating potential finite-time singularities in ideal flows.
%
\citet{Campolina2018_Chaotic} developed a model identical to the Euler equations by imitating the calculus on a 3D logarithmic lattice.
This model for ideal flows elucidates the emergence of singularities as a manifestation of a chaotic attractor in a renormalized dynamical system.
Their results implied that the direct numerical simulation (DNS) with the available resolution is inadequate for the analysis of singularity formation for the Euler equation.
%
By employing a level-set representation for the vorticity field, \citet{Constantin2001_Local, Constantin2001_NS} and \citet{Deng2005_A} established global existence theorems for a wide range of initial values and revealed the geometric structures of plausible blowup scenario, for the 3D Euler equations and the 3D Lagrangian averaged Euler equations.

In the present study, we apply the spherical Clebsch mapping~\citep{Kuznetsov1980_On} to develop the spin Euler equation, which is equivalent to the original Euler equation, but its form has an inherent Lagrangian nature for tracking vortex surfaces.
%
We study the Lagrangian dynamics and regularity of the spin Euler equation with the DNS of various inviscid vortical flows, and derive a new non-blowup condition for the spin Euler equation.
%
%

The outline of the present paper is as follows.
Section~\ref{sec:theo_spin_Euler} introduces the spin Euler equation and derives the non-blowup condition.
Section~\ref{sec:DNS_cases} describes numerical set-ups and methods.
Section~\ref{sec:evolution} elucidates Lagrangian dynamics of ideal flows and assesses the non-blowup criterion.
Some conclusions are drawn in \S\,\ref{sec:conclusion}.

\section{Theoretical framework of the spin Euler equation}\label{sec:theo_spin_Euler}
\subsection{Introduction of the spin Euler equation}
The 3D incompressible Euler equation is
\begin{equation}\label{eq:EulerEq}
    \p_t\vec{u} + \vec{u}\cdot\bn\vec{u}  = - \bn p
\end{equation}
with $\bn\cdot\vec{u} = 0$, where $\vec{u}$ is the velocity and $p$ the pressure.

By applying the spherical Clebsch mapping~\citep{Kuznetsov1980_On}, \eqref{eq:EulerEq} is transformed into a Lagrangian form
\begin{equation}\label{eq:Lagrangian_s}
    \p_t\vec{s} + \vec{u}\cdot\bn\vec{s} = \vec{0}.
\end{equation}
Here, the spin vector $\vec{s}\in\mathbb{S}^2$~\citep{Chern2016_Schrodinger, Chern2017_Fluid} is linked with a two-component wave function $\vec{\psi}=[\psi_1,\psi_2]\T\in\mathbb{S}^3$ by the Hopf fibration~\citep{Hopf1931_Uber}
\begin{equation}\label{eq:s}
    \vec{s} = (a^2+b^2-c^2-d^2, 2(bc-ad), 2(ac+bd)),
\end{equation}
with $\psi_1=a+\ii b$, $\psi_2=c+\ii d$ and the imaginary unit $\ii$.
Then, the velocity and vorticity $\vec{\omega}\equiv\bn\times\vec{u}$ can be re-expressed by $\vec{u} = a\bn b - b\bn a + c\bn d - d\bn c$ and
\begin{equation}\label{eq:vor_s}
    \vec{\omega} = \frac{1}{4}\varepsilon_{ijk}s_i\bn s_j\times \bn s_k,
\end{equation}
respectively, where $\varepsilon_{ijk}$ is the Levi--Civita symbol.
Note that even though an arbitrary $\vec{\omega}$ in \eqref{eq:vor_s} may have no globally smooth spin vector due to vorticity nulls~\citep{Graham2000_Clebsch} and unclosed vortex lines, a useful approximation can be obtained using the regularizer~\citep{Chern2017_Inside} and the Poincar\'e recurrence theorem~\citep{Poincare1890_Sur}.

We consider the quaternion form of the two-component wave function $\vec{\psi}=a+\vec{i}b+\vec{j}c+\vec{k}d$, where $a$, $b$, $c$ and $d$ are real-valued functions and $\{\vec{i},\vec{j},\vec{k}\}$ are the basis vectors of the imaginary part of the quaternion.
The velocity and spin vector are then given by $\vec{u}=(\bn\widebar{\vec{\psi}}\vec{i\psi} - \widebar{\vec{\psi}}\vec{i}\bn\vec{\psi})/2$ and $\vec{s}=\widebar{\vec{\psi}}\vec{i\psi}$, respectively, where $\widebar{\vec{\psi}}$ denotes the quaternion conjugate of $\vec{\psi}$.
Then, we derive
\begin{align}
    \vec{u}\cdot\bn\vec{s}
    =\ & \frac{1}{2}\left(\bn\widebar{\vec{\psi}}\vec{i\psi}\cdot \bn\widebar{\vec{\psi}}\vec{i\psi} - |\bn\vec{\psi}|^2 + |\bn\vec{\psi}|^2 - \widebar{\vec{\psi}}\vec{i}\bn\vec{\psi}\cdot \widebar{\vec{\psi}}\vec{i}\bn\vec{\psi}\right)
    \notag \\
    =\ & \frac{1}{2}\left(\vec{s}\bn\widebar{\vec{\psi}}\cdot\vec{i}\bn\vec{\psi} -\bn\widebar{\vec{\psi}}\cdot\vec{i}\bn\vec{\psi}\vec{s}\right) = \vec{s}\times\vec{m},
    \label{eq:s_times_m}
\end{align}
where $\vec{m} \equiv \bn\widebar{\vec{\psi}}\cdot\vec{i}\bn\vec{\psi}$ is a pure quaternion (i.e.~a vector in $\mathbb{R}^3$) and can be expanded as $\vec{m}=\vec{i}m_1+\vec{j}m_2+\vec{k}m_3$ with
\begin{equation}\label{eq:m}
    \begin{cases}
        m_1 = |\bn a|^2 + |\bn b|^2 - |\bn c|^2 - |\bn d|^2, \\
        m_2 = 2(\bn b\cdot\bn c - \bn a\cdot\bn d), \\
        m_3 = 2(\bn a\cdot\bn c + \bn b\cdot\bn d).
    \end{cases}
\end{equation}
Thus, we rewrite $\vec{u}\cdot\bn\vec{s}= \vec{s}\times\vec{m}$ with an effective field $\vec{m}=(m_1,m_2,m_3)$.

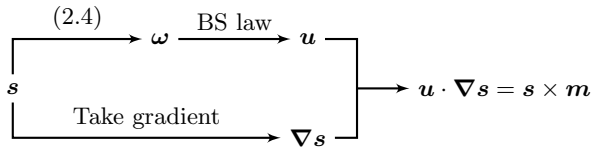
\begin{figure}
    \centering
    \tikzstyle{line1} = [draw, thick]
    \tikzstyle{line2} = [draw, -latex', thick]

    \begin{tikzpicture}[node distance = 6em]
        \node [] (vel) at (0, 0) {$\vec{u}$};
        \node [left of=vel] (vor) {$\vec{\omega}$};
        \node [] (s) at ($(vor)+(-6em,-2em)$) {$\vec{s}$};
        \node [below of=vel, node distance=4em] (grads) {$\bn\vec{s}$};
        \node [] at ($(vel)+(8em,-2em)$) (ucdotgrads) {$\vec{u}\cdot\bn\vec{s}=\vec{s}\times\vec{m}$};
        \path [line2] (s) -- ($(s)+(0em,2em)$) -- node [above] {\eqref{eq:vor_s}} (vor);
        \path [line2] (vor) -- node [above] {BS law} (vel);
        \path [line2] (s) -- ($(s)+(0em,-2em)$) -- node [above] {Take gradient} (grads);
        \path [line1] (vel) -- ($(vel)+(2em,0em)$) -- ($(grads)+(2em,0em)$) -- (grads);
        \path [line2] ($(vel)+(2em,-2em)$) -- (ucdotgrads);
    \end{tikzpicture}
    \caption{Schematic for computing $\vec{s}\times\vec{m}$.}
    \label{fig:cal_s_times_m}
\end{figure}

In general, $\vec{m}$ cannot be represented solely in terms of $\vec{s}$, because the Hopf mapping is irreversible.
Nevertheless, using the generalised Biot--Savart (BS) law, we can compute $\vec{s}\times\vec{m}$ at any point in $\mathbb{R}^3$ from the distribution of $\vec{s}$ under various boundary conditions, as illustrated in figure~\ref{fig:cal_s_times_m}.

Substituting \eqref{eq:s_times_m} into \eqref{eq:Lagrangian_s}, we obtain the spin Euler equation
\begin{equation}\label{eq:spin_Euler}
    \p_t\vec{s} + \vec{s}\times\vec{m} = \vec{0}.
\end{equation}
It is equivalent to the original incompressible Euler equation \eqref{eq:EulerEq}.
In contrast to \eqref{eq:Lagrangian_s}, \eqref{eq:spin_Euler} characterises the evolution of $\vec{s}$ by its precession about $\vec{m}$ rather than the convection with $\vec{u}$.
The spin Euler equation \eqref{eq:spin_Euler} can be more suitable to study fluid dynamics from a Lagrangian perspective than its original form~\eqref{eq:EulerEq}, because the isosurfaces of $s_i,~i=1,2,3$ are vortex surfaces consisting of vortex lines \citep{Yang2010_On, Yang2011_Evolution, Yang2023_Applications}.
From the Helmholtz theorem, the surfaces are material surfaces for all $t\ge 0$ in Euler flows.

Therefore, solving the spin Euler equation \eqref{eq:spin_Euler} is similar to a vortex method \citep{Yang2021_Clebsch, Nabizadeh2021_Covector, Xiong2022_A} for simulating ideal flows.
Since the primary variable $\vec{s}$ of \eqref{eq:spin_Euler} is a unit vector, the simulation of the spin Euler equation could avoid issues of numerical blowup.

In particular, the spin Euler equation contains the inherent Lagrangian vortex dynamics via level sets of $s_i$ (i.e.~vortex surfaces).
This can facilitate the regularity analysis of the Euler equation, similar to the level set representation of $\vec{\omega}$ \citep{Constantin2001_Local, Constantin2001_NS, Deng2005_A}.

It is also interesting that the spin Euler equation~\eqref{eq:spin_Euler} can be considered as the Landau--Lifshitz equation
\begin{equation}\label{eq:LLeq}
    \p_t\vec{s} + \vec{s}\times\vec{H}_{\mathrm{eff}} = \vec{0}
\end{equation}
without a damping term~\citep{Landau1935_On}, which is used to analyse magnetodynamic processes in magnetic materials, and is generally referred to as the Heisenberg model \citep{Lakshmanan1990_Planar, Porsezian1991_On, Kamppeter2001_Topological}.
The Landau--Lifshitz equation plays an important role in elucidating magnetisation dynamics, analogous to the role of the NS equation in fluid dynamics.

The mean spin of electrons, i.e.~the magnetisation (or spin vector) $\vec{s}$ at the macroscopic scale, determines the unit volume magnetic dipole moment in magnetic crystals.
A continuous function $\vec{s}(\vec{x},t)$ describes the macroscopic magnetisation dynamics in the limit of vanishing lattice partition size, if the angle between the spin vectors of neighbouring lattice atoms in a crystal is sufficiently small \citep{Heisenberg1928_Zur}.
This resembles the continuum assumption in fluid mechanics, but unlike isotropic fluids, most crystal structures are anisotropic.

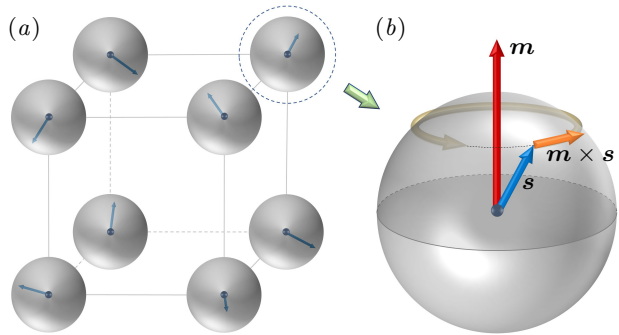
\begin{figure}
    \centering
    \begin{overpic}[width=0.6\textwidth]{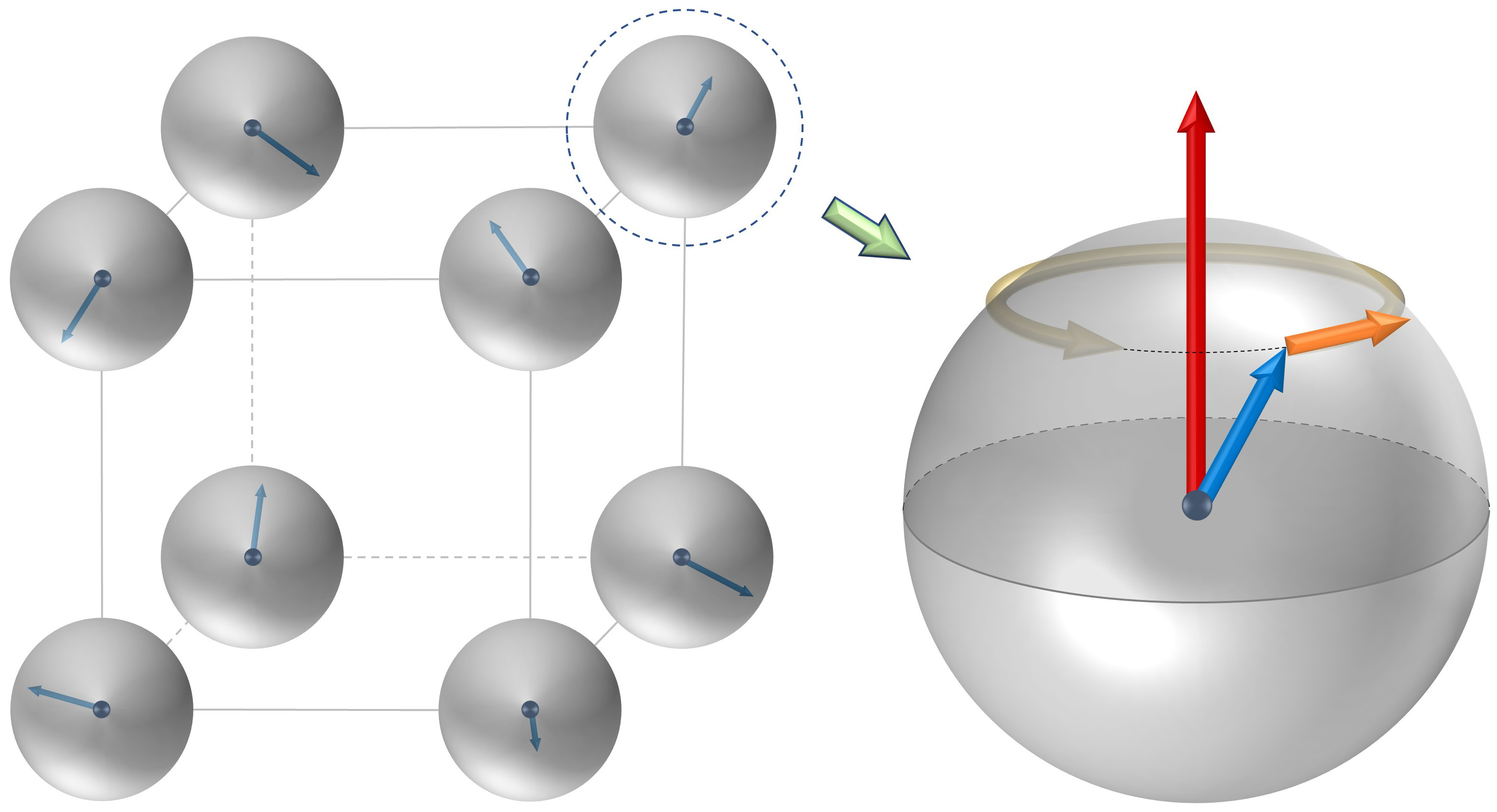}
        \put(0,49) {\fontsize{9}{1}\selectfont (\textit{a})}
        \put(60,49) {\fontsize{9}{1}\selectfont (\textit{b})}
        \put(82,46) {\fontsize{9}{1}\selectfont $\vec{m}$}
        \put(84,24) {\fontsize{9}{1}\selectfont $\vec{s}$}
        \put(88,28.5) {\fontsize{9}{1}\selectfont $\vec{m}\times\vec{s}$}
    \end{overpic}
    \caption{(\textit{a}) Sketch of a $2^3$ lattice of spin vectors, where the orientation of $\vec{s}$ fully characterizes the ideal flow by \eqref{eq:vor_s}.
    (\textit{b}) At each grid point, the magnetisation (or spin) $\vec{s}$ in \eqref{eq:s} precesses around the effective magnetic field $\vec{H}_{\mathrm{eff}}=\vec{m}$ in \eqref{eq:m}.}
    \label{fig:sphere}
\end{figure}

In \eqref{eq:LLeq}, $\vec{H}_{\mathrm{eff}}$ is the effective magnetic field, corresponding to (minus) the $L^2$-derivative of the magnetic energy of the material with respect to $\vec{s}$.
This implies a deep connection between the ideal flow and magnetic material.
The spin Euler equation has $\vec{H}_{\mathrm{eff}}=\vec{m}$, where the magnetic energy of a material is replaced by the total kinetic energy of a fluid.  
Therefore, the ideal flow might be physically interpreted as a specific magnetic material by \eqref{eq:spin_Euler}.
As sketched in figure~\ref{fig:sphere}, the magnetisation (or spin) $\vec{s}$ in \eqref{eq:s} at each point in space precesses around the effective magnetic field $\vec{m}$ in \eqref{eq:m}.
%

The similarity and difference between the spin Euler equation \eqref{eq:spin_Euler} and the isotropic Heisenberg spin system widely used in modelling magnetic crystals are discussed in Appendix~\ref{app:relation_to_Heisenberg}.
%


\subsection{Non-blowup condition of the spin Euler equation}
Next, we discuss the regularity of the spin Euler equation.
By taking the double inner-product of $\bn\vec{s}$ and the gradient of \eqref{eq:spin_Euler}, and multiplying $p|\bn\vec{s}|^{2(p-1)}$, we obtain
\begin{equation}\label{eq:dgrads2dt}
    \p_t|\bn\vec{s}|^{2p}
    = 2p|\bn\vec{s}|^{2(p-1)}\bn\vec{s}:(\bn\vec{m}\times\vec{s}),
\end{equation}
where $p\ge 1$ is a constant, and the vector product of the second-order tensor $\bn\vec{m}$ and the vector $\vec{s}$ is defined as $\bn\vec{m}\times\vec{s}\equiv \varepsilon_{ljk}\p_im_js_k\vec{e}_i\vec{e}_l$ with the basis $\{\vec{e}_1,\vec{e}_2,\vec{e}_3\}$.
Using the identity $\vec{s}\cdot\rmDelta\vec{s}=-|\bn\vec{s}|^2$ for the unit vector $\vec{s}$, we have
\begin{equation}\label{eq:ineq_grads_Laplacians}
    |\bn\vec{s}|^2 \le |\rmDelta\vec{s}|.
\end{equation}
Integrating \eqref{eq:dgrads2dt} over an unbounded domain $\mathcal{D}$ or a domain bounded by solid wall boundaries and using \eqref{eq:ineq_grads_Laplacians} yield
\begin{align}
    \left|\p_t\int_{\mathcal{D}}|\bn\vec{s}|^{2p} \dif V\right|
    &= 2p\left|\int_{\mathcal{D}}|\bn\vec{s}|^{2(p-1)}\bn\vec{s}:(\bn\vec{m}\times\vec{s})\dif V\right|
    \notag \\
    &\le  2p\norm{\rmDelta\vec{s}}_{\infty}^{p-1}\left|\int_{\mathcal{D}}\left[\bn\cdot(\bn\vec{s}\cdot(\vec{m}\times\vec{s})) + \vec{m}\cdot(\bn\cdot(\bn\vec{s}\times\vec{s})) \right]\dif V\right|
    \notag \\
    &= 2p\norm{\rmDelta\vec{s}}_{\infty}^{p-1}\left|\oiint_{\p\mathcal{D}}\frac{\p\vec{s}}{\p n}\cdot(\vec{m}\times\vec{s})\dif S \right.
    \notag \\
    &\hspace{7em}+ \left.\int_{\mathcal{D}}[\rmDelta\vec{s}\cdot(\vec{s}\times\vec{m}) + \bn\vec{s}:(\bn\vec{s}\times\vec{m})]\dif V\right|
    \notag \\
    &= 2p\norm{\rmDelta\vec{s}}_{\infty}^{p-1}\left|\int_{\mathcal{D}}\vec{m}\cdot(\rmDelta\vec{s}\times\vec{s})\dif V\right|,
    \label{eq:dnorm2gradsdt}
\end{align}
where the Neumann boundary condition $\p\vec{s}/\p n|_{\p\mathcal{D}}=\vec{0}$ is imposed on the solid wall boundary.
Applying the H\"older inequality to \eqref{eq:dnorm2gradsdt} yields
\begin{equation}\label{eq:ineq_pgrads2pt_mLp_LaplaciansLq}
    \p_t\norm{\bn\vec{s}}_{{2p}}^{2p}
    \le 2p\norm{\vec{m}}_{1} \norm{\rmDelta\vec{s}}_{\infty}^p.
\end{equation}
Here, the $L^p$-norm of a function $f$ is defined as
\begin{equation}
    \norm{f}_{p} \equiv
    \begin{dcases}
    \left(\int_{\mathcal{D}}|f|^p\dif V \right)^{1/p}, & 1\le p<\infty, \\
    \mathrm{esssup}|f|, & p=\infty.
    \end{dcases}
\end{equation}

We estimate the upper bound of $\norm{\vec{\omega}}_{p}$ in terms of $\norm{\bn\vec{s}}_{{2p}}$.
From \eqref{eq:vor_s}, using $|\vec{s}|=1$ and basic inequalities, 
we obtain
\begin{align}
    |\vec{\omega}|
    &\le \frac{1}{2}(|\bn s_1||\bn s_2| + |\bn s_2||\bn s_3| + |\bn s_3||\bn s_1|)
    \notag \\
    &\le \frac{1}{2}\left[\frac{1}{2}(|\bn s_1|^2+|\bn s_2|^2) + \frac{1}{2}(|\bn s_2|^2+|\bn s_3|^2) + \frac{1}{2}(|\bn s_3|^2+|\bn s_1|^2)\right]
    \notag \\
    &= \frac{1}{2}|\bn\vec{s}|^2,
    \label{eq:ieq_vor_grads}
\end{align}
so that
\begin{equation}\label{eq:ineq_vorLp_gradsL2p}
    \|\vec{\omega}\|_{p}
    \le \frac{1}{2}\|\bn\vec{s}\|_{{2p}}^2.
\end{equation}

Then, we estimate the upper bound of $\norm{\vec{m}}_{1}$ in terms of $\norm{\bn\vec{s}}_{{2}}$ and $\norm{\rmDelta\vec{s}}_{\infty}$.
Substituting \eqref{eq:ieq_vor_grads} into \eqref{eq:m}, we have
\begin{align}
    |\vec{m}|^2
    =\ & |\bn a|^4 + |\bn b|^4 + |\bn c|^4 + |\bn d|^4 + 2|\bn a|^2|\bn b|^2 + 2|\bn c|^2|\bn d|^2
    \notag \\
    &- 2|\bn a|^2|\bn c|^2 - 2|\bn a|^2|\bn d|^2 - 2|\bn b|^2|\bn c|^2 - 2|\bn b|^2|\bn d|^2
    \notag \\
    &+ 4[(\bn b\cdot\bn c)^2 + (\bn a\cdot\bn d)^2 - 2(\bn b\cdot\bn c)(\bn a\cdot\bn d)]
    \notag \\
    &+ 4[(\bn a\cdot\bn c)^2 + (\bn b\cdot\bn d)^2 + 2(\bn a\cdot\bn c)(\bn b\cdot\bn d)]
    \notag \\
    \le \ & (|\bn a|^2 + |\bn b|^2 + |\bn c|^2 + |\bn d|^2)^2
    \notag \\
    &+ 8[(\bn a\cdot\bn c)(\bn b\cdot\bn d) - (\bn a\cdot\bn d)(\bn b\cdot\bn c)]
    \notag \\
    =\ & |\bn\vec{\psi}|^4 + 8(\bn a\times\bn b)\cdot(\bn c\times\bn d)
    \notag \\
    =\ & \left(\frac{1}{4}|\bn\vec{s}|^2 + |\vec{u}|^2 \right)^2 + 4\left(\frac{|\vec{\omega}|^2}{4} - |\bn a\times\bn b|^2 - |\bn c\times\bn d|^2 \right)
    \notag \\
    \le \ & \frac{1}{8}|\bn\vec{s}|^4 + 2|\vec{u}|^4 + |\vec{\omega}|^2
    \notag \\
    \le \ & \frac{3}{8}|\bn\vec{s}|^4 + 2|\vec{u}|^4,
\end{align}
so that
\begin{align}
    |\vec{m}| &\le \left(\frac{3}{8}|\bn\vec{s}|^4 + 2|\vec{u}|^4\right)^{1/2}
    \notag \\
    &= \left[\left(\frac{\sqrt{6}}{4}|\bn\vec{s}|^2 + \sqrt{2}|\vec{u}|^2 \right)^2 - \sqrt{3}|\bn\vec{s}|^2|\vec{u}|^2 \right]^{1/2}
    \notag \\
    & \le \frac{\sqrt{6}}{4}|\bn\vec{s}|^2 + \sqrt{2}|\vec{u}|^2.
    \label{eq:ineq_m}
\end{align}
Integrating \eqref{eq:ineq_m} over the domain $\mathcal{D}$ yields
\begin{equation}\label{eq:ieq_m_grads}
    \|\vec{m}\|_{1}
    \le \frac{\sqrt{6}}{4}\|\bn\vec{s}\|_{{2}}^2 + \sqrt{2}\norm{\vec{u}}_{2}^2.
\end{equation}
By virtue of the Sobolev--Poincar\'e inequality~\citep{Moffatt1992_Helicity},
\begin{equation}\label{eq:ineq_u_vor}
    \norm{\vec{u}}_{2}^2
    \le C_\omega\norm{\vec{\omega}}_{2}^2
\end{equation}
holds, with a positive constant $C_\omega$ independent of $\vec{u}$.
Combining \eqref{eq:ineq_grads_Laplacians}, \eqref{eq:ineq_vorLp_gradsL2p}, \eqref{eq:ieq_m_grads} and \eqref{eq:ineq_u_vor}, we derive
\begin{align}
    \norm{\vec{m}}_{1}
    &\le \frac{\sqrt{6}}{4}\|\bn\vec{s}\|_{{2}}^2 + \sqrt{2}C_\omega\norm{\vec{\omega}}_{2}^2
    \notag \\
    &\le \frac{\sqrt{6}}{4}\|\bn\vec{s}\|_{{2}}^2 + \frac{\sqrt{2}}{4}C_\omega\norm{\bn\vec{s}}_{4}^4
    \notag \\
    &\le \frac{\sqrt{6}}{4}\|\bn\vec{s}\|_{{2}}^2 + \frac{\sqrt{2}}{4}C_\omega\norm{\bn\vec{s}}_{2}^2\norm{\rmDelta\vec{s}}_{\infty}
    \notag \\
    &= \frac{1}{4}\norm{\bn\vec{s}}_{2}^2\left(\sqrt{6} + \sqrt{2}C_\omega\norm{\rmDelta\vec{s}}_{\infty} \right).
    \label{eq:ieq_m_grads_1}
\end{align}

Finally, substituting \eqref{eq:ieq_m_grads_1} into \eqref{eq:ineq_pgrads2pt_mLp_LaplaciansLq} and applying the H\"older inequality, we have
\begin{equation}\label{eq:ineq_pgradsL2pt_1}
    \frac{1}{\norm{\bn\vec{s}}_{{2p}}^2}\p_t\norm{\bn\vec{s}}_{{2p}}^{2p}
    \le \frac{p}{2}\mu^{(p-1)/p}(\mathcal{D})\left(\sqrt{6} + \sqrt{2}C_\omega\norm{\rmDelta\vec{s}}_{\infty} \right)\norm{\rmDelta\vec{s}}_{\infty}^p,
\end{equation}
where $\mu(\mathcal{D})$ is the finite measure of domain $\mathcal{D}$.
Integrating \eqref{eq:ineq_pgradsL2pt_1} over time yields
\begin{equation}\label{eq:ineq_grads_L2}
    \norm{\bn\vec{s}(\cdot,t)}_{2}^2
    \le C_s\exp\left(\frac{1}{2}\int_0^t\left(\sqrt{6} + \sqrt{2}C_\omega\norm{\rmDelta\vec{s}(\cdot,\tau)}_{\infty} \right)\norm{\rmDelta\vec{s}(\cdot,\tau)}_{\infty}\dif\tau \right)
\end{equation}
for $p=1$, with a constant $C_s = \exp\left(\norm{\bn\vec{s}(\cdot,0)}_{2}^2 \right)$, and
\begin{equation}\label{eq:ineq_grads_L2p}
    \norm{\bn\vec{s}(\cdot,t)}_{{2p}}^{2(p-1)}
    \le \frac{p-1}{2}\mu^{(p-1)/p}(\mathcal{D})\int_0^t \left(\sqrt{6} + \sqrt{2}C_\omega\norm{\rmDelta\vec{s}(\cdot,\tau)}_{\infty} \right)\norm{\rmDelta\vec{s}(\cdot,\tau)}_{\infty}^p\dif\tau
\end{equation}
for $p>1$.

As the H\"older inequality leads to
\begin{equation}
    \int_0^t \norm{\rmDelta\vec{s}(\cdot,\tau)}_{\infty}^p\dif\tau
    \le t^{1/(p+1)}\left(\int_0^t \norm{\rmDelta\vec{s}(\cdot,\tau)}_{\infty}^{p+1}\dif\tau\right)^{p/(p+1)},
\end{equation}
we have
\begin{align}
    &\int_0^t\left(\sqrt{6} + \sqrt{2}C_\omega\norm{\rmDelta\vec{s}(\cdot,\tau)}_{\infty} \right)\norm{\rmDelta\vec{s}(\cdot,\tau)}_{\infty}^p\dif\tau
    \notag \\
    \le&\ \sqrt{6}t^{1/(p+1)}\left(\int_0^t \norm{\rmDelta\vec{s}(\cdot,\tau)}_{\infty}^{p+1}\dif\tau\right)^{p/(p+1)}
    + \sqrt{2}C_\omega\int_0^t \norm{\rmDelta\vec{s}(\cdot,\tau)}_{\infty}^{p+1}\dif\tau
    \notag \\
    \le& C_p(t)\int_0^t \norm{\rmDelta\vec{s}(\cdot,\tau)}_{\infty}^{p+1}\dif\tau, \quad \forall p\ge 1,
    \label{eq:ineq_Laplcians2_Laplcians}
\end{align}
where $C_p(t)=\sqrt{6}t^{1/(p+1)}+\sqrt{2}C_\omega$ is a finite coefficient when $t<\infty$.
Using \eqref{eq:ineq_vorLp_gradsL2p}, \eqref{eq:ineq_grads_L2} and \eqref{eq:ineq_Laplcians2_Laplcians}, the $L^1$-norm of the vorticity can be estimated by
\begin{equation}\label{eq:ineq_vor_L1}
    \norm{\vec{\omega}}_{1}
    \le \frac{C_s}{2}\exp\left(\frac{C_1(t)}{2}\int_0^t \norm{\rmDelta\vec{s}(\cdot,\tau)}_{\infty}^{2}\dif\tau \right).
\end{equation}
Moreover, using \eqref{eq:ineq_vorLp_gradsL2p}, \eqref{eq:ineq_grads_L2p} and \eqref{eq:ineq_Laplcians2_Laplcians}, the $L^{p}$-norm ($p>1$) of the vorticity can be estimated by
\begin{equation}\label{eq:ineq_vor_Lp}
    \norm{\vec{\omega}}_{p}
    \le \frac{1}{2}\left(\frac{p-1}{2}\mu^{(p-1)/p}(\mathcal{D})C_p(t)\int_0^t\norm{\rmDelta\vec{s}(\cdot,\tau)}_{\infty}^{p+1}\dif\tau \right)^{1/(p-1)}.
\end{equation}
From \eqref{eq:ineq_vor_L1} and \eqref{eq:ineq_vor_Lp}, we obtain a sufficient condition for bounded $\norm{\vec{\omega}}_{p}$ as
\begin{equation}\label{eq:non-blowup_condition_LaplacianS}
    \int_0^t\norm{\rmDelta\vec{s}(\cdot,\tau)}_{\infty}^{p+1}\dif\tau < \infty, \quad \forall p\ge 1.
\end{equation}
Note that it is straightforward to deduce the special case of \eqref{eq:non-blowup_condition_LaplacianS} with $p=2$ from \eqref{eq:ieq_vor_grads} using the BKM theorem~\citep{Beale1984_Remarks}.

In summery, we obtain a non-blowup condition \eqref{eq:non-blowup_condition_LaplacianS} of the spin Euler equation (which is equivalent to the original Euler equation), which guarantees a bounded $\norm{\vec{\omega}}_{p}$.
It implies that, if the solution loses regularity beyond a certain time, the Laplacian of the spin vector must grow unbounded.
The transport equation and the estimation of the norm of $\rmDelta\vec{s}$ are further discussed in Appendix~\ref{app:Lap_s}.


\section{Simulation overview}\label{sec:DNS_cases}
We conduct the DNS of three ideal flows with different initial conditions in a periodic cube of side $2\pi$ on $N^3$ (up to $1536^3$) uniform grid points, by solving the spin Euler equation
\begin{equation}\label{eq:spin_Euler_num}
    \begin{dcases}
        \frac{\p\vec{s}}{\p t} = \mathcal{F}^{-1}\left[\frac{1}{|\vec{\kappa}|^2}\ii\mathcal{F}\left(\frac{1}{4}\varepsilon_{ijk}s_i\bn s_j\times \bn s_k \right) \times\vec{\kappa} \right]\cdot\bn\vec{s}\\
        \vec{s}(\vec{x},t=0) = \vec{s}_0(\vec{x})
    \end{dcases}
\end{equation}
with the pseudo-spectral method.
Here, $\vec{\kappa}$ denotes the wavenumber vector, $\vec{s}_0$ is a smooth initial condition, and $\mathcal{F}$ is the Fourier transform operator with its inverse form $\mathcal{F}^{-1}$.
The high-order Fourier smoothing method~\citep{Hou2007_Computing} is used to suppress the Gibbs phenomenon.
The temporal evolution is integrated using an explicit second-order Runge--Kutta scheme with adaptive time steps in physical space.
The time step is selected to ensure that the Courant--Friedrichs--Lewy number is smaller than 0.3 for numerical stability and accuracy.

We consider two types of initial conditions.
For the first type, the initial vorticity is concentrated in a thin closed vortex tube, such as the trefoil knot~\citep{Yao2021_Dynamics, Zhao2021_Helicity, Zhao2021_Direct} and Hopf link~\citep{Aref1991_Linking, Kivotides2021_Helicity, Yao2022_Helicity}.
Under the self-induced velocity, such vortex tubes can be gradually stretched, twisted and flattened, and form nearly singular vortical structures.

We use the rational map~\citep{Kedia2016_Weaving, Tao2021_Construction} to construct smooth $\vec{s}_0$.
A small twist is applied to the vortex tube by setting $P=\alpha$~\citep{Tao2021_Construction}, and $Q=\alpha^3+\beta^2$ and $Q=\alpha^2+\beta^2$ are chosen for the trefoil knot and the Hopf link, respectively.
Here, $(P,Q)$ are a pair of complex polynomial functions and $(\alpha,\beta)$ is a mapping of the coordinate system from the Euclidean space $\mathbb{R}^3$ to the two-component complex space $\mathbb{C}^2$.
The function pair $(P,Q)$ is normalised and subjected to a divergence-free projection, yielding in a two-component wave function $\vec{\psi}_0=[\psi_{1,0},\psi_{2,0}]\T$ that matches the initial field.
The initial spin vector $\vec{s}_0$ and vorticity $\vec{\omega}_0$ are then obtained from \eqref{eq:s} and \eqref{eq:vor_s}, respectively.
Additionally, we re-scale the time as $t^*=t/(L_0^2/\varGamma)$ with the initial mean length $L_0=2\sqrt{2}\pi^{3/2}/\norm{\bn\vec{s}(\cdot,0)}_{2}$ and the circulation $\varGamma$.
The trefoil knot has $L_0=0.749$ and $\varGamma=5.05$, and the Hopf link has $L_0=0.773$ and $\varGamma=5.24$.

The second type is a modified Taylor--Green (MTG) initial condition~\citep{Meng2023_Quantum}, with smooth
\begin{equation}
  \vec{s}_0=(\cos x\cos y\cos z,\sqrt{1-\cos^2x\cos^2y\cos^2z}\cos 2z,\sqrt{1-\cos^2x\cos^2y\cos^2z}\sin 2z)
\end{equation}
and
\begin{equation}
  \vec{\omega}_0=(\cos x\sin y\cos z,-\sin x\cos y\cos z,0).
\end{equation}
The re-scaling time is $t^*=t$.
This highly symmetric MTG flow would not exhibit a finite-time singularity,
and this non-blowup case is used to validate the criterion in \eqref{eq:non-blowup_condition_LaplacianS}.
The parameters for all cases are listed in table~\ref{tab:cases}.

\begin{table}
    \centering
    \setlength{\tabcolsep}{4mm}{
    \begin{tabular}{cccccccc}
        Cases & $N^3$ & $\norm{\vec{u}_0}_2^2$ & $\mathcal{H}_0$ & $L_0$ & $\varGamma$ & $t^*/t$ & $T_\mathcal{R}$  \\
        Trefoil knot & up to $1536^3$ & $72.4$ & $79.0$ & $0.749$ & $5.05$ & $9.00$ & $2.39$ \\
        Hopf link & up to $1536^3$ & $68.9$ & $79.0$ & $0.773$ & $5.24$ & $8.77$ & $2.61$ \\
        MTG & up to $1024^3$ & $20.7$ & $0$ & -- & -- & $1$ & $3.67$ \\
    \end{tabular}}
    \caption{DNS cases and parameters.}
    \label{tab:cases}
\end{table}

\begin{figure}
    \centering
    \includegraphics[width=\textwidth]{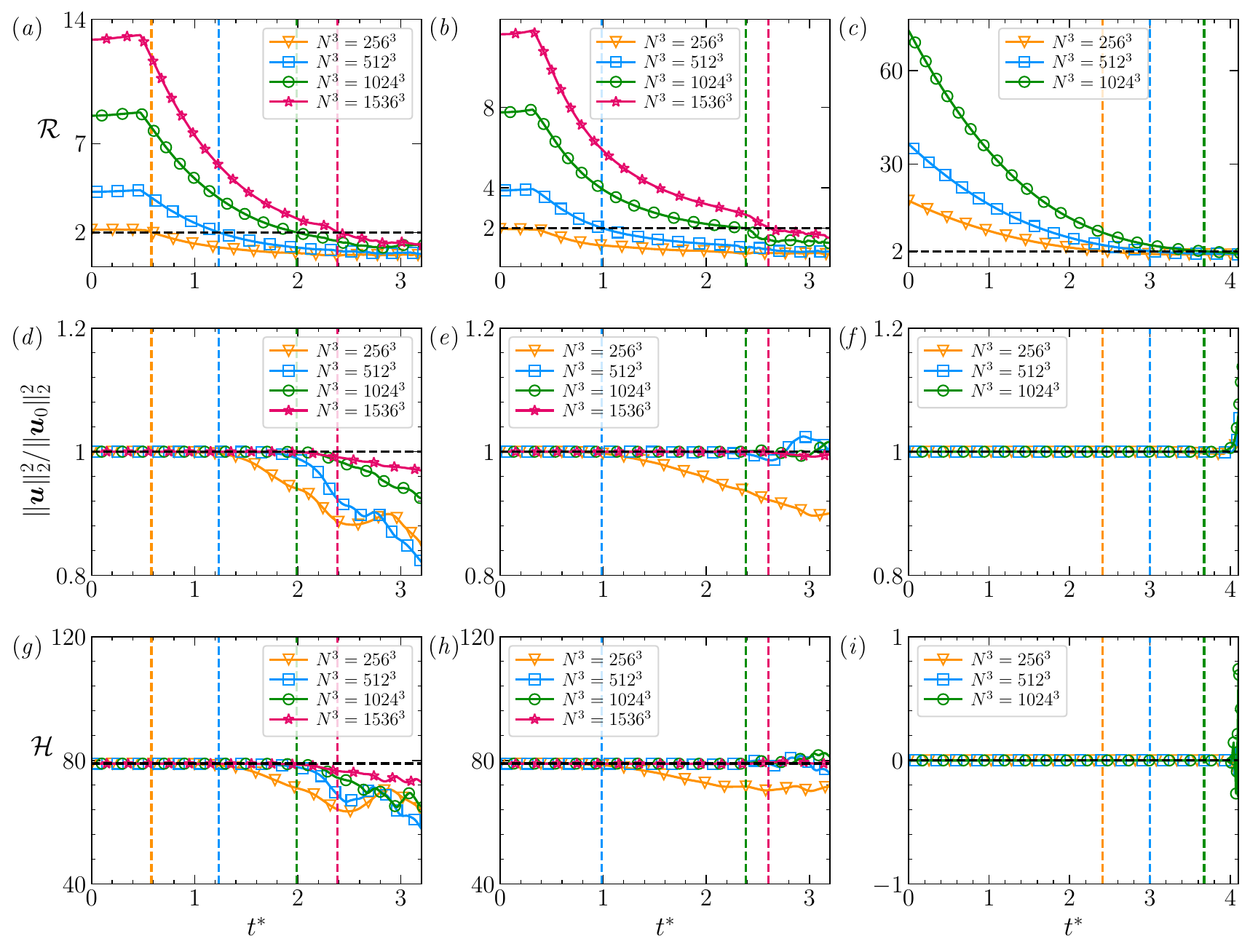}
    \caption{Evolution of (\textit{a}--\textit{c}) $\mathcal{R}(t^*)\equiv1/(h\norm{\bn\vec{s}(\cdot,t^*)}_{\infty})$, (\textit{d}--\textit{f}) $\norm{\vec{u}}_{2}^2/\norm{\vec{u_0}}_{2}^2$ and (\textit{g}--\textit{i}) the helicity for the (\textit{a},\textit{d},\textit{g}) trefoil knot, (\textit{b},\textit{e},\textit{h}) Hopf link and (\textit{c},\textit{f},\textit{i}) MTG flow, with different grid resolutions. The vertical dashed lines with different colours mark the time with $\mathcal{R}=2$ for each resolutions. The simulations are well-resolved on the left of the dashed lines.}
    \label{fig:resolution_energy_helicity}
\end{figure}

To evaluate the numerical resolution, we define $\mathcal{R}(t^*)\equiv1/(h\norm{\bn\vec{s}(\cdot,t^*)}_{\infty})$, the ratio of the minimum resolved scale to the grid spacing $h$.
A finer resolution has a larger $\mathcal{R}$.
The evolution of $\mathcal{R}$ for the three initial conditions is shown in figure~\ref{fig:resolution_energy_helicity}(\textit{a}--\textit{c}).
Our numerical tests suggest that $\mathcal{R}\ge 2$ can be the criterion for well resolving the smallest scale of \eqref{eq:spin_Euler_num}.
The largest numbers of grid points in the simulation are $N^3=1536^3$ for the trefoil knot and the Hopf link, and $N^3=1024^3$ for the MTG flow.
Based on the criterion, the largest time $T_\mathcal{R} \equiv t^*|_{\mathcal{R}=2}$ of the simulation with the satisfactory resolution is given in table~\ref{tab:cases} for each case.

Additionally, the resolution can be assessed by the conservation of the total energy $\norm{\vec{u}}_{2}^2$ and the helicity $\mathcal{H}=\int_{\mathcal{D}}\vec{u}\cdot\vec{\omega}\dif V$ \citep{Moreau1961_Constantes, Moffatt1969_The, Meng2023_Evolution}, which are two invariants of the Euler equations.
The energy loss is less than 1\textperthousand~for $\mathcal{R}\ge 2$ in figures~\ref{fig:resolution_energy_helicity}(\textit{d}--\textit{f}), and the helicity is also well conserved in figures~\ref{fig:resolution_energy_helicity}(\textit{g}--\textit{i}).

\section{Evolution of the spin Euler equation}\label{sec:evolution}

\subsection{Evolution of vortex surfaces}
\begin{figure}
    \centering
    \vspace{1em}
    \begin{overpic}[width=0.7\textwidth]{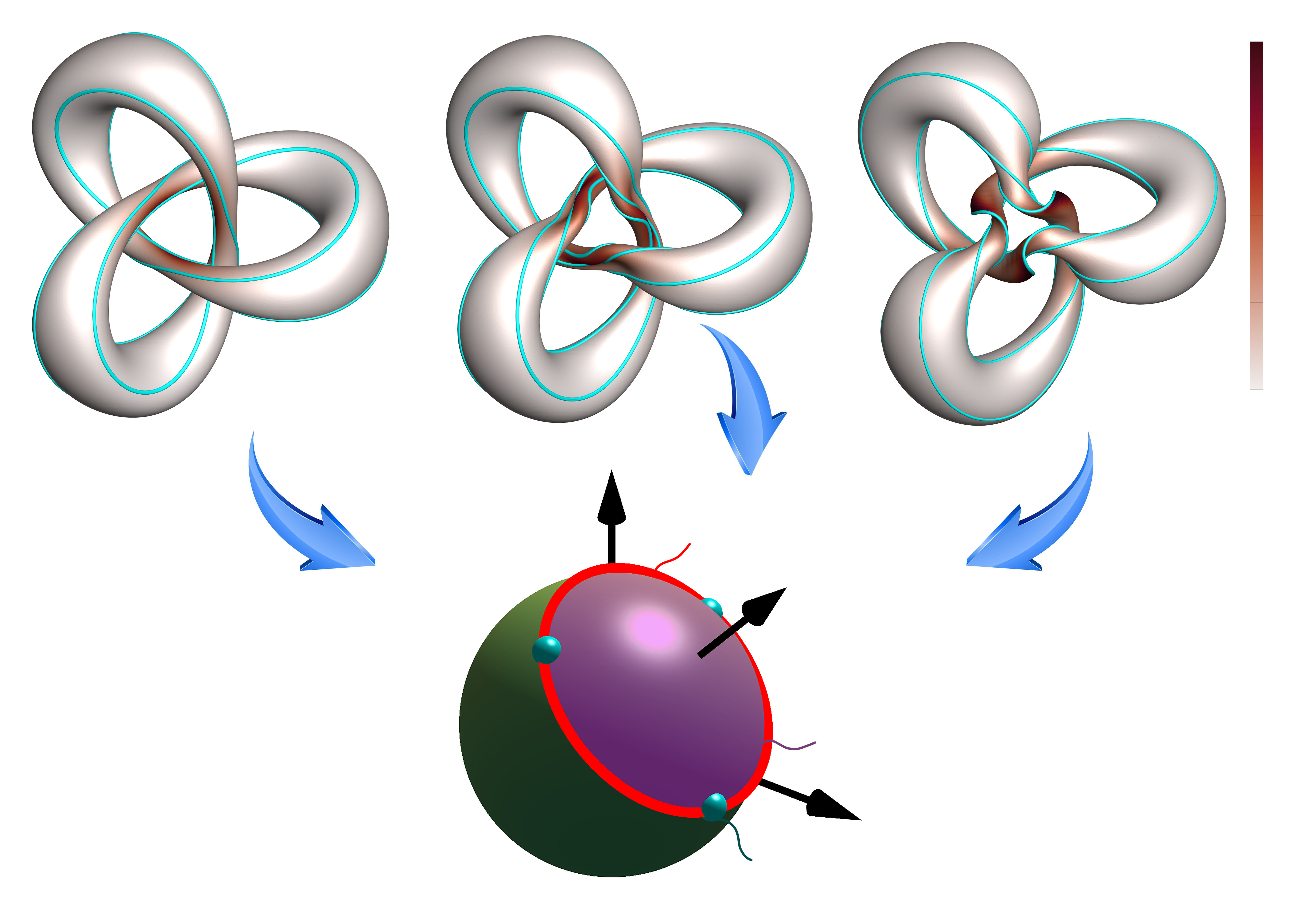}
        \put(1,67) {\fontsize{9}{1}\selectfont (\textit{a}) $t^*=0$}
        \put(33,67) {\fontsize{9}{1}\selectfont (\textit{b}) $t^*=0.9$}
        \put(64.5,67) {\fontsize{9}{1}\selectfont (\textit{c}) $t^*=1.8$}
        \put(31,27) {\fontsize{9}{1}\selectfont (\textit{d})}
        \put(93,68.5) {\fontsize{9}{1}\selectfont $|\vec{\omega}|$}
        \put(96.5,64.7) {\fontsize{8}{1}\selectfont $100$}
        \put(96.5,57.825) {\fontsize{8}{1}\selectfont $75$}
        \put(96.5,51.25) {\fontsize{8}{1}\selectfont $50$}
        \put(96.5,44.675) {\fontsize{8}{1}\selectfont $25$}
        \put(96.5,38.3) {\fontsize{8}{1}\selectfont $0$}
        \put(60.5,24.5) {\fontsize{9}{1}\selectfont $s_1$}
        \put(66,5.5) {\fontsize{9}{1}\selectfont $s_2$}
        \put(45,34) {\fontsize{9}{1}\selectfont $s_3$}
        \put(52,28) {\fontsize{8}{1}\selectfont \red{vortex surface}}
        \put(62.5,11.5) {\fontsize{8}{1}\selectfont \textcolor [rgb]{1,0,1}{vortex tube}}
        \put(57.5,1) {\fontsize{8}{1}\selectfont \textcolor [rgb]{0,0.5,0.5}{vortex line}}
    \end{overpic}
    \caption{(\textit{a}--\textit{c}) Evolution of the isosurface of $s_1=0.5$ (vortex surfaces) colour-coded by $|\vec{\omega}|$ for the trefoil knot at $t^*=0$, 0.9 and 1.8 in the top view. Some vortex lines (cyan) are integrated and plotted on the isosurfaces. (\textit{d}) The vortex surface for $s_1=0.5$, the region enclosed by this vortex surface, and the three cyan vortex lines are map to the red circle, the purple spherical cap within the red circle and the three cyan points on the Bloch sphere $\mathbb{S}^2$, respectively.}
    \label{fig:T23_vortexsurface}
\end{figure}

The DNS of the spin Euler equation \eqref{eq:spin_Euler} is carried out to investigate Lagrangian dynamics of ideal flows listed in table~\ref{tab:cases}, and to validate the non-blowup criterion \eqref{eq:non-blowup_condition_LaplacianS}.
To illustrate the Lagrangian vortex dynamics, figure~\ref{fig:T23_vortexsurface} shows the top view of the isosurface of $s_1=0.5$ (i.e.~vortex surface) for the trefoil knot at $t^*=0$, 0.9 and 1.8.
Note that isosurfaces of $s_2$ and $s_3$ can show similar structures \citep{Tao2021_Construction}, and the isosurfaces of $|\vec{\omega}|$ (not shown) fail to capture the complete vortex tube as visualised by $s_1$ \citep[as discussed in][]{Xiong2019_Identifying, Shen2023_Role} .

Near the three crossings of the initial vortex knot, adjacent parts of the vortex tube are nearly orthogonal.
Driven by the self-induced velocity with the BS law, the vortex tube and vortex lines are stretched and twisted. The adjacent parts of the vortex knot approach each other, and they are progressively flattened and rolled up, instead of undergoing the vortex reconnection in viscous flows~\citep{Yao2022_Vortex}.
%
%
The regions with large vorticity magnitude $|\vec{\omega}|$ are rapidly stretched into spiral sheets with strong twist.

\begin{figure}
    \centering
    \vspace{1em}
    \begin{overpic}[width=0.7\textwidth]{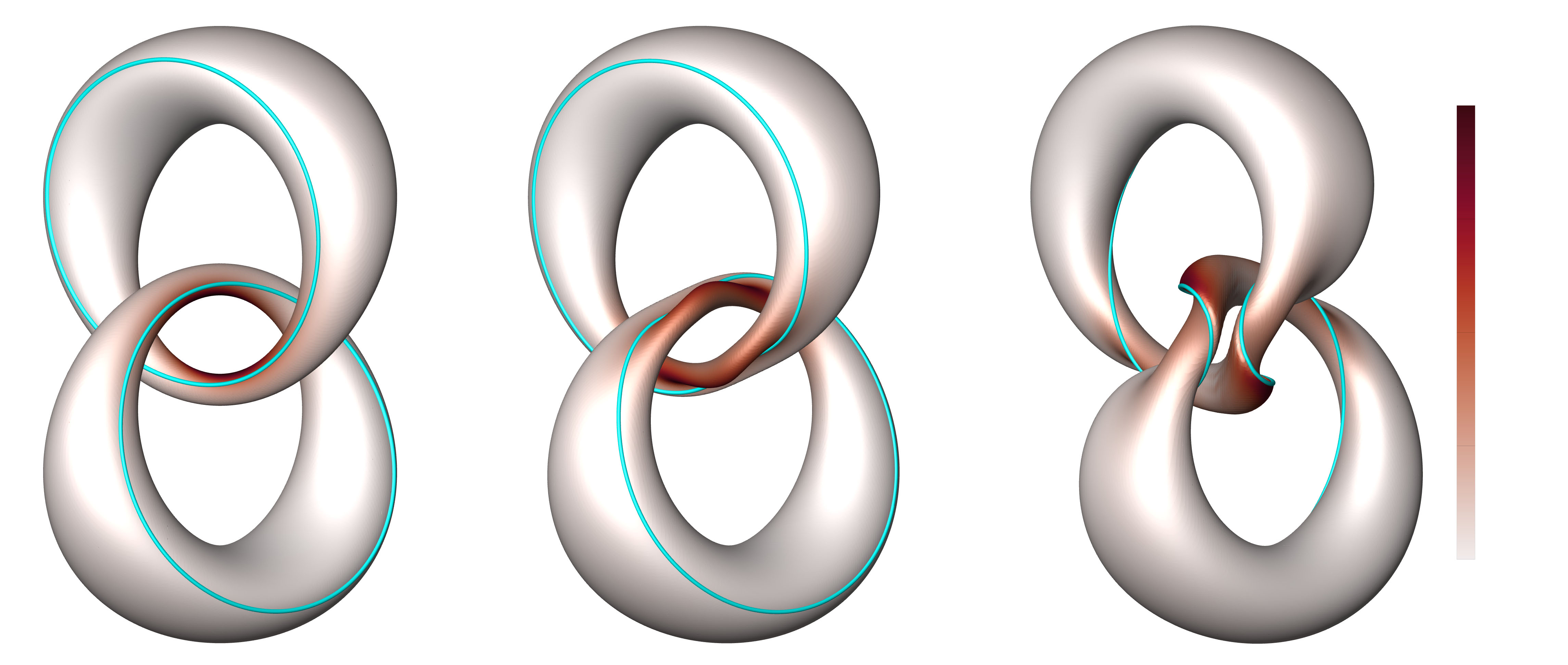}
        \put(1,41.4) {\fontsize{9}{1}\selectfont (\textit{a}) $t^*=0$}
        \put(32.5,41.4) {\fontsize{9}{1}\selectfont (\textit{b}) $t^*=0.88$}
        \put(64,41.4) {\fontsize{9}{1}\selectfont (\textit{c}) $t^*=2.19$}
        \put(90.8,38) {\fontsize{9}{1}\selectfont $|\vec{\omega}|$}
        \put(95,34.8) {\fontsize{8}{1}\selectfont $100$}
        \put(95,27.5) {\fontsize{8}{1}\selectfont $75$}
        \put(95,20.4) {\fontsize{8}{1}\selectfont $50$}
        \put(95,13.1) {\fontsize{8}{1}\selectfont $25$}
        \put(95,5.9) {\fontsize{8}{1}\selectfont $0$}
    \end{overpic}
    \caption{Evolution of the isosurface of $s_1=0.5$ (vortex surface) colour-coded by $|\vec{\omega}|$ for the Hopf link at $t^*=0$, 0.88 and 2.19 in the top view. Some vortex lines (cyan) are integrated and plotted on the isosurfaces.}
    \label{fig:T22_vortexsurface}
\end{figure}
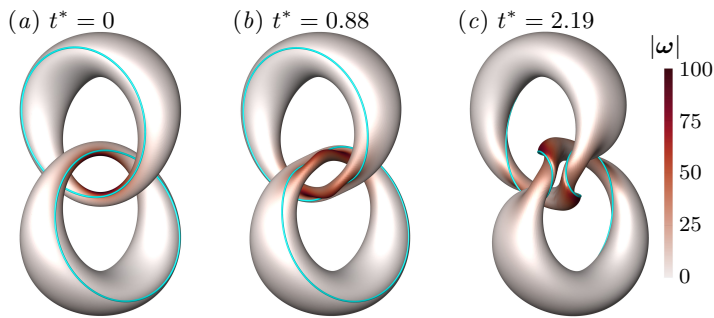

In figure~\ref{fig:T23_vortexsurface}, the evolving vortex surfaces and lines preserve their initial mapping to the red circle and cyan points on $\mathbb{S}^2$, respectively, due to the Lagrangian nature of the spin Euler equation.
Namely, the vortex topology is preserved in ideal flows.
In addition, figure~\ref{fig:T22_vortexsurface} shows the top view of the isosurface of $s_1=0.5$ for the Hopf link at $t^*=0$, 0.88, and 2.19.
The structural evolution is similar to that of the trefoil knot. 
%

\begin{figure}
    \centering
    \begin{overpic}[width=0.7\textwidth]{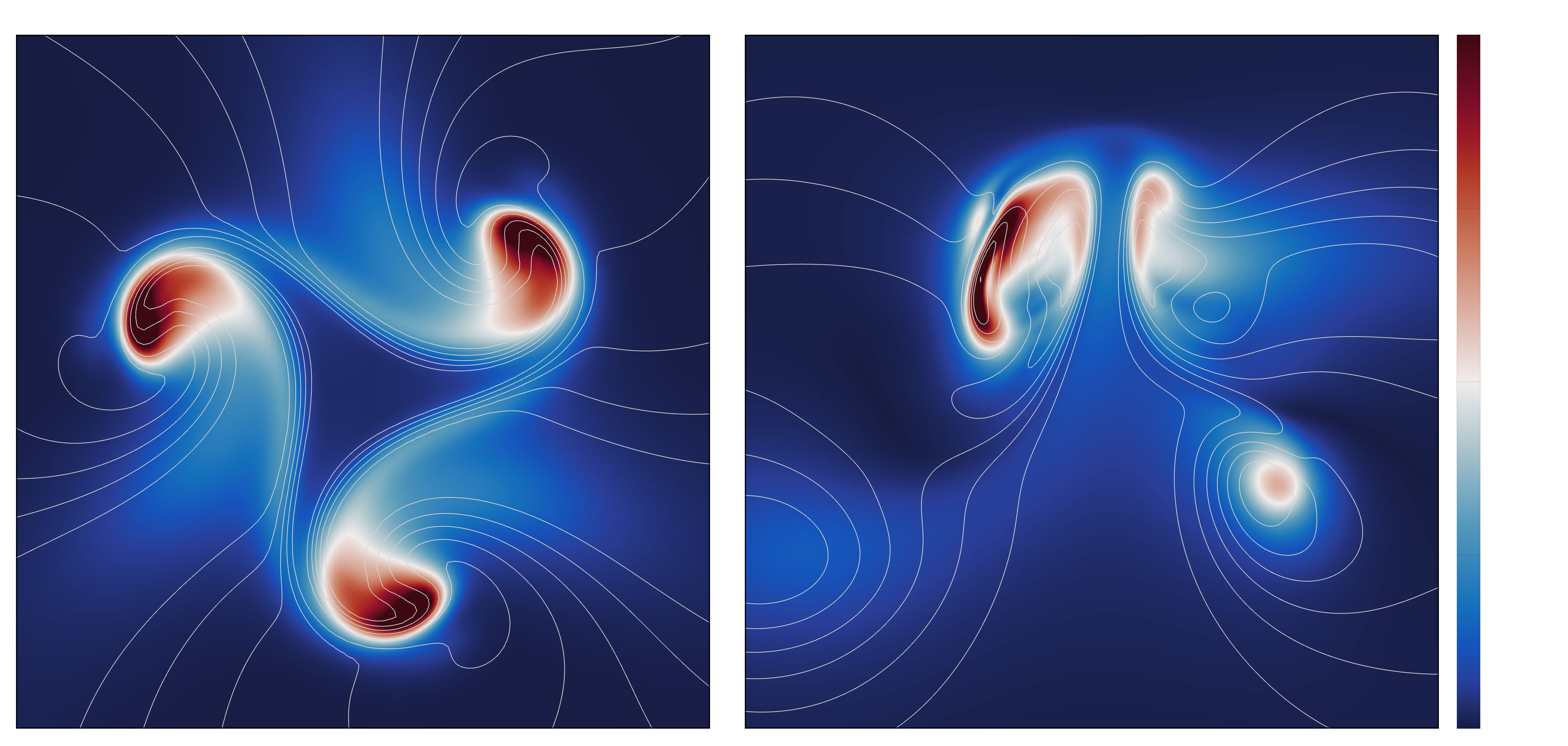}
        \put(95,44.3) {\fontsize{8}{1}\selectfont $100$}
        \put(95,33.3) {\fontsize{8}{1}\selectfont $75$}
        \put(95,22.3) {\fontsize{8}{1}\selectfont $50$}
        \put(95,11.3) {\fontsize{8}{1}\selectfont $25$}
        \put(95,0.3) {\fontsize{8}{1}\selectfont $0$}
        \put(1,42) {\fontsize{9}{1}\selectfont \white{(\textit{a}) $x$--$y$ plane}}
        \put(47.5,42) {\fontsize{9}{1}\selectfont \white{(\textit{b}) $y$--$z$ plane}}
    \end{overpic}
    \caption{Contour of the vorticity magnitude and contour lines (white) of $s_1$ for the trefoil knot on the (\textit{a}) $x$--$y$ plane at $z=2.55$ and (\textit{b}) $y$--$z$ plane at $x=3.24$, at $t^*=1.8$. The planes intersecting the point with the largest $|\vec{\omega}|$ contain the most intense swirling motion in the flow.}
    \label{fig:T23_vor_slice_line}
\end{figure}
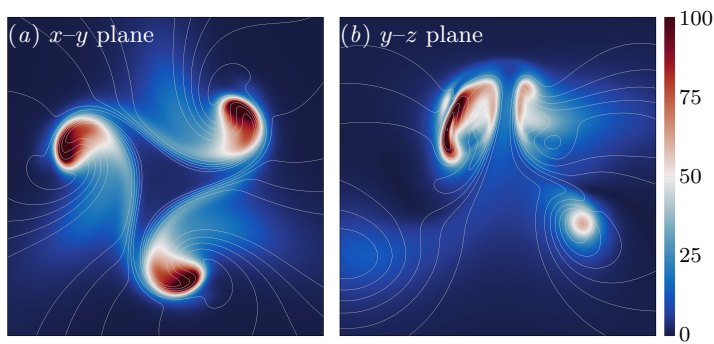

Figure~\ref{fig:T23_vor_slice_line} plots the contour of $|\vec{\omega}|$ on the $x$--$y$ plane at $z=2.55$ and on the $y$--$z$ plane at $x=3.24$ for the trefoil knot, along with the contour lines of $s_1$. 
These planes intersect the point with the largest $|\vec{\omega}|$, so their contours show the most intense swirling motion.
In figure~\ref{fig:T23_vor_slice_line}(\textit{b}), `vorticity pancakes'~\citep{Brachet1992_Numerical} form in the regions of large $|\vec{\omega}|$ among highly stretched and curved vortex surfaces.
These structures appear when the vortex surfaces approach each other and undergo strong deformation.
The formation of the high-vorticity region within sheet-like structures was observed in the collapse of vortex pairs \citep[e.g.][]{Pumir1990_Collapsing, Kerr1993_Evidence} and TG and KP flows \citep{Yang2010_On}.
Furthermore, we observe the energy spectra with the $k^{-3}$ scaling (not shown) in the evolution of the trefoil knot and Hopf link, consistent with the result for the collision of two Lamb dipoles in \citet{Orlandi2012_Vortex}.

\begin{figure}
    \centering
    \begin{overpic}[width=\textwidth]{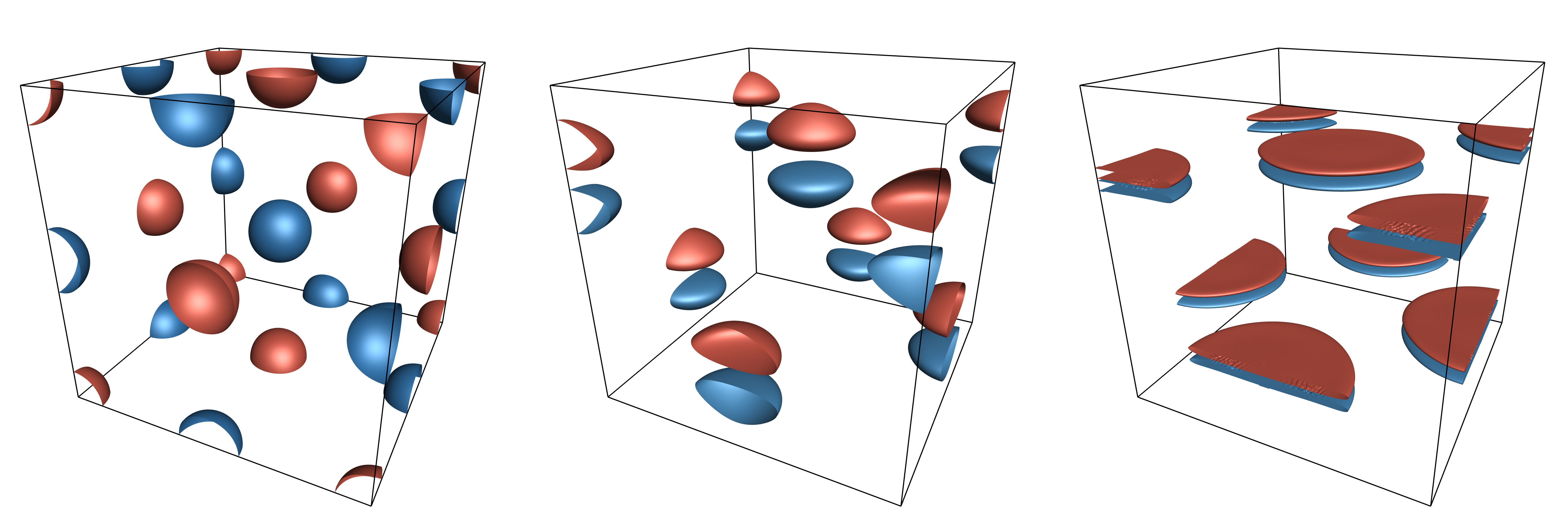}
        \put(0,31) {\fontsize{9}{1}\selectfont (\textit{a}) $t^*=0$}
        \put(34,31) {\fontsize{9}{1}\selectfont (\textit{b}) $t^*=2$}
        \put(68,31) {\fontsize{9}{1}\selectfont (\textit{c}) $t^*=4$}
    \end{overpic}
    \caption{Evolution of the isosurfaces of $s_1=0.8$ (red) and $s_1=-0.8$ (blue) in the MTG flow.}
    \label{fig:MTG_vortexsurface}
\end{figure}

In the highly symmetric MTG flow, a finite-time singularity may not exhibit according to the theoretical analysis \citep{Constantin1996_Geometric}.
Figure~\ref{fig:MTG_vortexsurface} plots the evolution of the isosurfaces of $s_1=0.8$ (red) and $s_1=-0.8$ (blue) for the MTG flow.
A pair of vortex blobs are compressed and flattened into pancakes.
%
Since the vortex surface is compressed in a quasi-2D configuration, preserving the smoothness  $\bn(\vec{\omega}/|\vec{\omega}|)$ of the vorticity direction \citep{Constantin1996_Geometric}, the MTG flow does not exhibit a finite time singularity, even though the vorticity grows rapidly \citep{Brachet1992_Numerical}.

The comparison of the two types of ideal flows implies that the Euler equation cannot form a singularity in a 2D process.
\citet{Constantin1996_Geometric} proved that if $\vec{u}$ remains uniformly bounded and $\vec{\omega}/|\vec{\omega}|$ stays $C^1$, then no singularity can occur.
In other words, the vorticity must change its direction very rapidly to form a potential singularity.
Note that if a singularity occurs at a vorticity null for all $t^*<t_b^*$ \citep[e.g.][]{Elgindi2021_Finite}, the vorticity direction becomes discontinuous at the time of singularity.
The MTG flow vortex lines near the vorticity nulls maintain a quasi-2D smooth shape, which contradicts the necessary blowup conditions.
Hence, the MTG flow does not exhibit finite-time singularities.
By contrast, the trefoil knot or Hopf link have large vortex-line curvature in figures~\ref{fig:T23_vortexsurface}(\textit{c}) and \ref{fig:T22_vortexsurface}(\textit{c}) with a rapid change in the vorticity direction. 

\subsection{Assessment of the non-blowup criterion}
\begin{figure*}
    \centering
    \includegraphics[width=\textwidth]{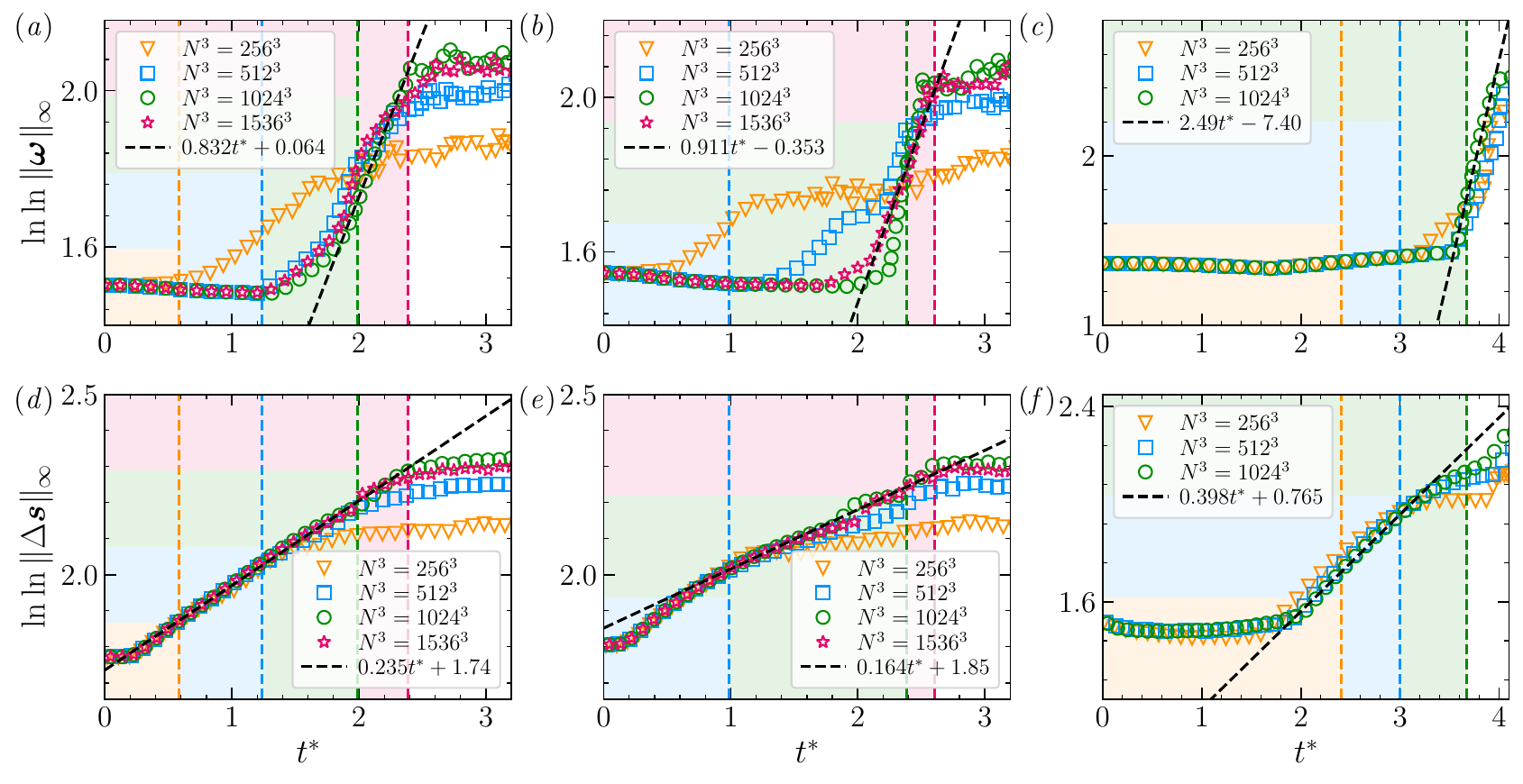}
    \caption{Evolution of (\textit{a}--\textit{c}) the maximum vorticity and (\textit{d}--\textit{f}) the maximum Laplacian spin vector for the (\textit{a},\textit{d}) trefoil knot, (\textit{b},\textit{e}) Hopf link and (\textit{c},\textit{f}) MTG flow, with different grid resolutions, respectively. Note that we plot $\ln\ln(50\norm{\vec{\omega}}_{\infty})$ for the MTG flow, the factor 50 is used to avoid complex values of the logarithm. The vertical dashed lines with different colours mark the time with $\mathcal{R}=2$ for each resolutions. The simulations are well-resolved on the left of the dashed lines (shaded in corresponding colours).}
    \label{fig:NormInf}
\end{figure*}

We apply \eqref{eq:non-blowup_condition_LaplacianS} to the three ideal flows to test the non-blowup criterion based on the spin Euler equation, and compare our criterion to the BKM criterion \citep{Beale1984_Remarks} by examining growth rates of the maximum vorticity and Laplacian spin vector.
Before $t^* = T_\mathcal{R}$, $\norm{\vec{\omega}}_{\infty}$ increases by a factor of about 16 for the trefoil knot and the Hopf link in figure~\ref{fig:NormInf}.
Both $\norm{\vec{\omega}}_{\infty}$ and $\norm{\rmDelta\vec{s}}_{\infty}$ exhibit the nearly double-exponential growth for the trefoil knot and Hopf link.
%
The double-exponential growth of $\norm{\vec{\omega}}_{\infty}$ is consistent with the results in \citet{Hou2007_Computing} and \citet{Kerr2013_Bounds}.
As the number of grid points increases (up to $1536^3$), the growth rate of $\norm{\rmDelta\vec{s}}_{\infty}$ appears to remain constant for the trefoil knot and Hopf link.

The present criterion has some advantages that $\ln\ln\norm{\rmDelta\vec{s}}_{\infty}$ grows more slowly than $\ln\ln\norm{\vec{\omega}}_{\infty}$ (4--6 times slower), and exhibits better convergence with the mesh resolution.
Therefore, $\rmDelta\vec{s}$ can be resolved more easily than $\vec{\omega}$ with the same numerical accuracy.
Moreover, the duration of the linear stage for $\ln\ln\norm{\rmDelta\vec{s}}_\infty$ exceeds that for $\ln\ln\norm{\vec{\omega}}_\infty$ by more than a factor of five. 
%

The highly symmetric MTG flow shows no evidence of a finite time singularity.
The profile of $\ln\ln\norm{\rmDelta\vec{s}}_\infty$ in figure~\ref{fig:NormInf}(\textit{f}) clearly bends downward before $t^* = T_\mathcal{R}$ with $N^3=1024^3$.
The growth rate of $\norm{\rmDelta\vec{s}}_\infty$ is weaker than double exponential, whereas the growth of $\norm{\vec{\omega}}_\infty$ remains double exponential in figure~\ref{fig:NormInf}(\textit{c}) when $t^*<T_{\mathcal{R}}$.
Therefore, the criterion based on $\rmDelta\vec{s}$ can effectively identify the flows that are unlikely to develop a finite-time singularity.


As the double-exponential growth is bounded in a finite time $t$, we have
\begin{equation}
    \int_0^t\norm{\rmDelta \vec{s}(\cdot,\tau)}_{\infty}^{p+1}\dif\tau
    \le \frac{(\ee^{c_1t}-1)\exp[(p+1)\ee^{c_1t+c_2}]}{c_1\ee^{c_1t}}
\end{equation}
with constants $c_1$ and $c_2$.
According to the non-blowup condition \eqref{eq:non-blowup_condition_LaplacianS}, the Euler equation can avoid singularity formation in finite time for the double-exponential growth of $\norm{\rmDelta\vec{s}}_\infty$.
%
%

\subsection{Difference of non-blowup criteria}
\begin{figure}
    \centering
    \includegraphics[width=0.8\textwidth]{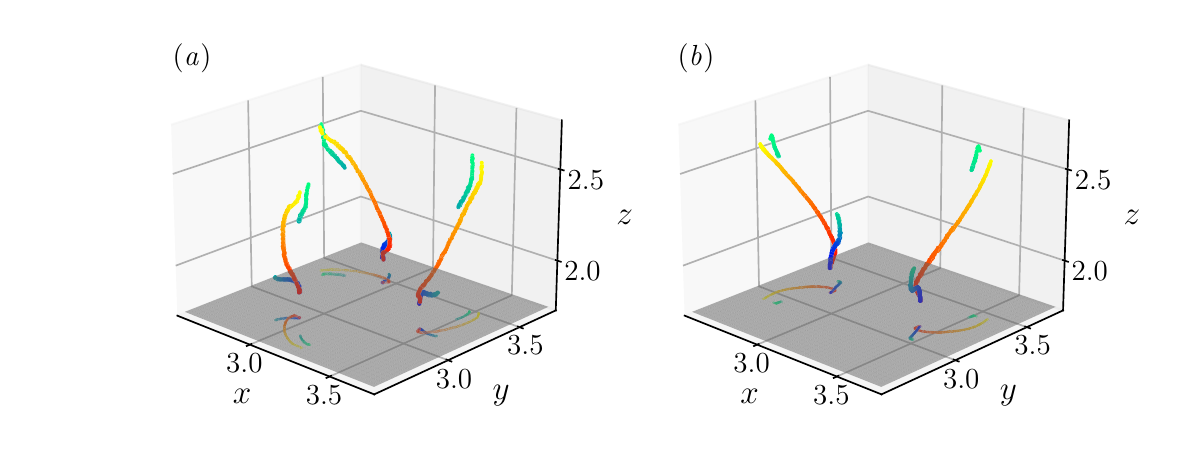}
    \caption{Three groups of trajectories of the maximum points of $|\vec{\omega}|$ (from blue to green) and $|\rmDelta\vec{s}|$ (from red to yellow) during $t^*\in[0,2.2]$ for the (\textit{a}) trefoil knot and (\textit{b}) Hopf link, along with their projections on the $x$--$y$ plane (shaded in gray). The two trajectories of maximum $|\vec{\omega}|$ and $|\rmDelta\vec{s}|$ in each group starts at the same location.}
    \label{fig:location_max}
\end{figure}

We highlight the major difference between \eqref{eq:non-blowup_condition_LaplacianS} and the BKM criterion, and explain why $\norm{\rmDelta\vec{s}}_{\infty}$ grows more slowly than $\norm{\vec{\omega}}_{\infty}$.
Figure~\ref{fig:location_max} plots the trajectories of $\arg\max|\vec{\omega}|$ and $\arg\max|\rmDelta\vec{s}|$ colour-coded by $t^*$ and their projections on the $x$--$y$ plane for the trefoil knot and Hopf link.
The trajectories of $\arg\max|\vec{\omega}|$ and $\arg\max|\rmDelta\vec{s}|$ starting from the same locations do not collapse, implying that the present criterion is distinct from the BKM criterion.

\begin{figure}
    \centering
    \includegraphics[width=0.9\textwidth]{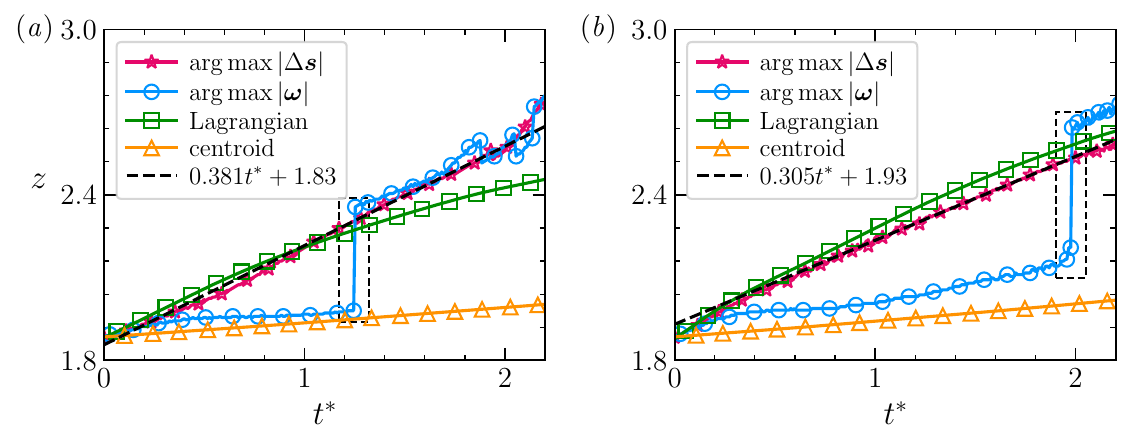}
    \caption{Evolution of the $z$-direction coordinates of the maximum points of $|\vec{\omega}|$ and $|\rmDelta\vec{s}|$, the Lagrangian tracing particle located at the maximum point at $t^*=0$ and the centroid position for the (\textit{a}) trefoil knot and (\textit{b}) Hopf link.}
    \label{fig:location_max_z}
\end{figure}

The continuous trajectory of $\arg\max|\rmDelta\vec{s}|$ is more tractable than the discontinuous one of $\arg\max|\vec{\omega}|$ in figure~\ref{fig:location_max}.
Figure~\ref{fig:location_max_z} illustrates the $z$-coordinates of the maximum $|\vec{\omega}|$ and $|\rmDelta\vec{s}|$, the Lagrangian trajectory of particles which locate at the position of the maximum values at $t^*=0$, and the evolution of the centroid positions $z_c = \int_{s_1\ge 0}z\dif V \big{/}\int_{s_1\ge 0}\dif V$ of the trefoil knot and Hopf link.
%
We find that $\arg\max|\rmDelta\vec{s}|$ remains continuous over time and moves at a constant speed in the $z$-direction in both flows, which is close to the Lagrangian velocity of the particle at the location of $\arg\max|\rmDelta\vec{s}|$ (or $\arg\max|\vec{\omega}|$) at $t^*=0$.
This implies that the maximum $|\rmDelta\vec{s}|$ could have some Lagrangian nature.

By contrast, $\arg\max|\vec{\omega}|$ exhibits a sharp jump at $t^*=1.26$ for the trefoil knot and $t^*=1.98$ for the Hopf link.
The speed of $\arg\max|\vec{\omega}|$ in the $z$-direction is close to that of the centroid of the vortex at early times. During the surge of $|\vec{\omega}|$, it jumps to a value close to $\arg\max|\rmDelta\vec{s}|$ (marked in dashed box in figure~\ref{fig:location_max_z}).

\begin{figure}
    \centering
    \includegraphics[width=0.8\textwidth]{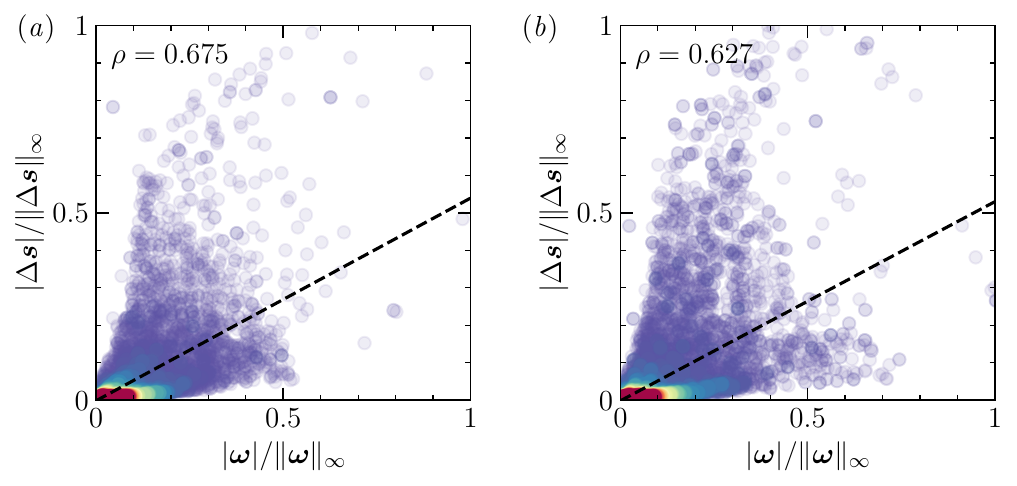}
    \caption{Scatter plots of $|\rmDelta\vec{s}|$ and $|\vec{\omega}|$ normalised with their maximum values for (\textit{a}) the trefoil knot at $t^*=1.8$ and (\textit{b}) the Hopf link at $t^*=2.19$, along with the correlation coefficients (marked at the upper left) and linear fits (dashed lines) of data points. The scattered point is coloured from purple to red by the density of data points.}
    \label{fig:scatter}
\end{figure}

We examine the correlation and distribution of the values of $|\rmDelta\vec{s}|$ and $|\vec{\omega}|$, normalised by their respective maxima, for the trefoil knot at $t^*=1.8$ and the Hopf link at $t^*=2.19$.
The scatter plots in figure~\ref{fig:scatter} shows a low positive correlation between $|\rmDelta\vec{s}|/\norm{\rmDelta\vec{s}}_{\infty}$ and $|\vec{\omega}|/\norm{\vec{\omega}}_{\infty}$, with correlation coefficients of $\rho=0.675$ for the trefoil knot and $\rho=0.627$ for the Hopf link.
Therefore, the criterion based on $\rmDelta\vec{s}$ in \eqref{eq:non-blowup_condition_LaplacianS} has a notable statistical difference from that based on $\vec{\omega}$.

\begin{figure}
    \centering
    \includegraphics[width=0.9\textwidth]{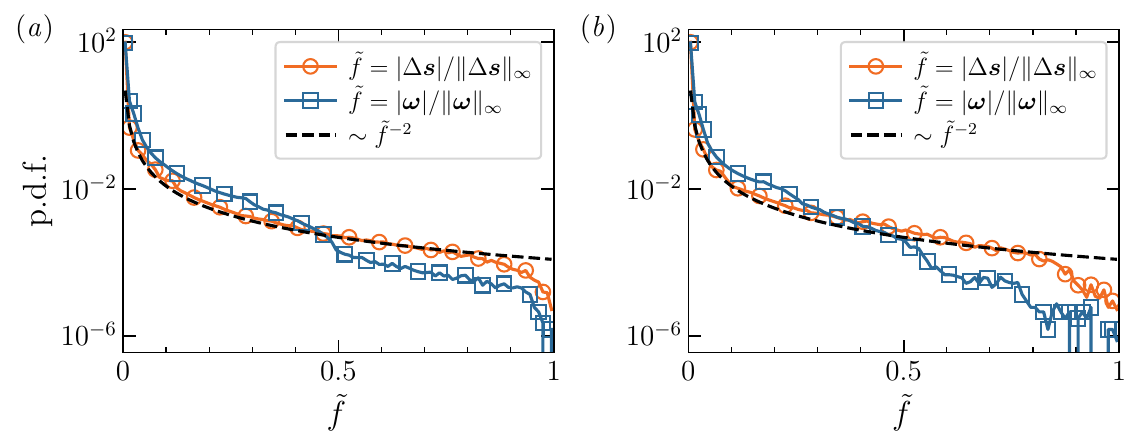}
    \caption{PDFs of normalised $|\rmDelta\vec{s}|$ and $|\vec{\omega}|$ for (\textit{a}) the trefoil knot at $t^*=1.8$ and (\textit{b}) the Hopf link at $t^*=2.19$.}
    \label{fig:pdf}
\end{figure}

The probability density function (PDF) of the normalised values of $|\rmDelta\vec{s}|$ and $|\vec{\omega}|$ for the trefoil knot at $t^*=1.8$ and the Hopf link at $t^*=2.19$ is shown in figure~\ref{fig:pdf}.
For both configurations, the PDF profiles of $|\rmDelta\vec{s}|/\norm{\rmDelta\vec{s}}_{\infty}$ are smoother than those of $|\vec{\omega}|/\norm{\vec{\omega}}_{\infty}$, and they obey a Pareto distribution \citep{Arnold2015_Pareto} with the $-2$ power law except for very large values, indicating that the extreme values at a few locations can dominate norms of $\rmDelta\vec{s}$ and $\vec{\omega}$.

\section{Conclusions}\label{sec:conclusion}
We develop a new framework for describing ideal flows using the spin Euler equation~\eqref{eq:spin_Euler}.
The spin Euler equation can be considered as a special Landau--Lifshitz equation with an effective magnetic field $\vec{H}_{\mathrm{eff}}=\vec{m}$ in \eqref{eq:m}, implying a possible connection between the ideal flow and magnetic material.

The spin Euler equation provides a feasible approach to study Lagrangian fluid dynamics, because the isosurfaces of a spin-vector component are vortex surfaces and material surfaces for all $t\ge 0$.
%
%
In particular, we derive a non-blowup condition \eqref{eq:non-blowup_condition_LaplacianS} for the spin Euler equation -- if the solution becomes singular at some finite time, then $|\rmDelta\vec{s}|$ must become unbounded.

We conduct the DNS of three ideal flows of the trefoil knot, Hopf link and MTG by solving \eqref{eq:spin_Euler_num} using the pseudo-spectral method, and compare the BKM criterion with the present one.
The evolution of the vortex surface (isosurface of $s_1$) illustrates that the regions with large $|\vec{\omega}|$ are rapidly stretched into spiral sheets for the trefoil knot and Hopf link.

For the trefoil knot and Hopf link, the double-exponential growth of $\norm{\rmDelta\vec{s}}_{\infty}$ is more pronounced than that of $\norm{\vec{\omega}}_{\infty}$,
%
and $\ln\ln\norm{\rmDelta\vec{s}}_{\infty}$ grows at a rate 4--6 times slower than $\ln\ln\norm{\vec{\omega}}_{\infty}$.
The duration of the double-exponential growth stage for $\norm{\rmDelta\vec{s}}_\infty$ exceeds that for $\norm{\vec{\omega}}_\infty$ by more than five times. 
%
%
According to the non-blowup condition \eqref{eq:non-blowup_condition_LaplacianS}, the Euler equation can avoid the singularity formation at finite time if the growth rate of $\norm{\rmDelta\vec{s}}_\infty$ is lower than double exponential at late times.

The highly symmetric MTG flow can avoid finite-time singularities due to the formation of quasi-2D vortex surfaces from theoretical analysis~\citep{Constantin1996_Geometric}.
The growth rate of $\norm{\rmDelta\vec{s}}_\infty$ is lower than double exponential at late times, whereas the growth rate of $\norm{\vec{\omega}}_\infty$ remains double exponential.
Thus, the present criterion based on $\rmDelta\vec{s}$ appears to be more sensitive than the BKM criterion based on $\vec{\omega}$ in detecting the flows that are incapable of producing finite-time singularities.

The present non-blowup criterion based on $|\rmDelta\vec{s}|$ is distinct from the BKM criterion based on $|\vec{\omega}|$.
By tracing the maxima of $|\rmDelta\vec{s}|$ and $|\vec{\omega}|$ for vortex knots and link, we find that the trajectory for $|\rmDelta\vec{s}|$ is continuous and consistent with the tracer particle, benefited from the Lagrangian nature of the spin Euler equation. In contrast, the trajectory for $|\vec{\omega}|$ with a large jump deviates from the Lagrangian trajectory.
Furthermore, $|\rmDelta\vec{s}|$ and $|\vec{\omega}|$ only has a low positive correlation coefficient.


In the future work, the bound estimate of $|\rmDelta\vec{s}|$ requires further refinement, and the duration in the simulation can be prolonged with more computational resources for examining longer growth behaviour of $\norm{\rmDelta\vec{s}}_\infty$.
Furthermore, the spin Euler equation can be recast as a nonlinear Schr\"odinger equation that is useful in quantum computing of fluid dynamics~\citep{Meng2023_Quantum}.
%

\section*{Acknowledgments}
The authors thank S. Xiong for the helpful discussion.
Numerical simulations were carried out on the TH-2A supercomputer in Guangzhou, China.

\section*{Funding}
This work has been supported by the National Natural Science Foundation of China (Grant Nos.~11925201 and 11988102), the National Key R\&D Program of China (Grant No.~2020YFE0204200), and the Xplore Prize.

\section*{Declaration of interests}
The authors report no conflict of interest.

\section*{Author contributions}
Y.Y. and Z.M. designed research. Z.M. preformed research. Y.Y. and Z.M. discussed the results and wrote the manuscript. All the authors have given approval for the manuscript.

\appendix

\section{Comparison of the spin Euler system and the isotropic Heisenberg spin system}\label{app:relation_to_Heisenberg}
We discuss the similarity and difference between the spin Euler system in \eqref{eq:spin_Euler} and the isotropic Heisenberg spin system in \eqref{eq:LLeq} with $\vec{H}_{\mathrm{eff}}=\rmDelta\vec{s}$.
After some algebra, we find
\begin{equation}
    \vec{m} = \frac{1}{2}\rmDelta\vec{s} + \vec{m}',
\end{equation}
where the term
\begin{equation}
    \vec{m}' = (-a\rmDelta a-b\rmDelta b+c\rmDelta c+d\rmDelta d,
    a\rmDelta d+d\rmDelta a-b\rmDelta c-c\rmDelta b,
    - a\rmDelta c-c\rmDelta a-b\rmDelta d-d\rmDelta b )
\end{equation}
highlights the difference between the spin Euler system and the Heisenberg spin system.
Projecting $\vec{m}'$ onto $\vec{s}$ yields
\begin{equation}
    \vec{s}\cdot\vec{m}'
    = -(a\rmDelta a+b\rmDelta b+c\rmDelta c+d\rmDelta d)
    = |\bn\vec{\psi}|^2
    \ge 0,
\end{equation}
i.e.~the angle between $\vec{m}'$ and $\vec{s}$ is acute or normal. 

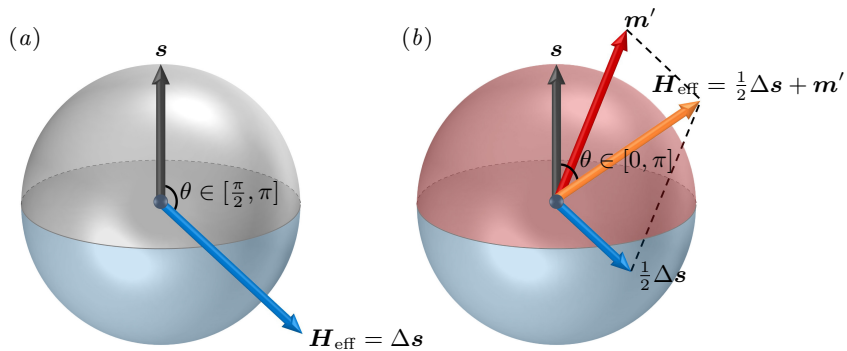
\begin{figure}
    \centering
    \vspace{1em}
    \begin{overpic}[width=0.7\textwidth]{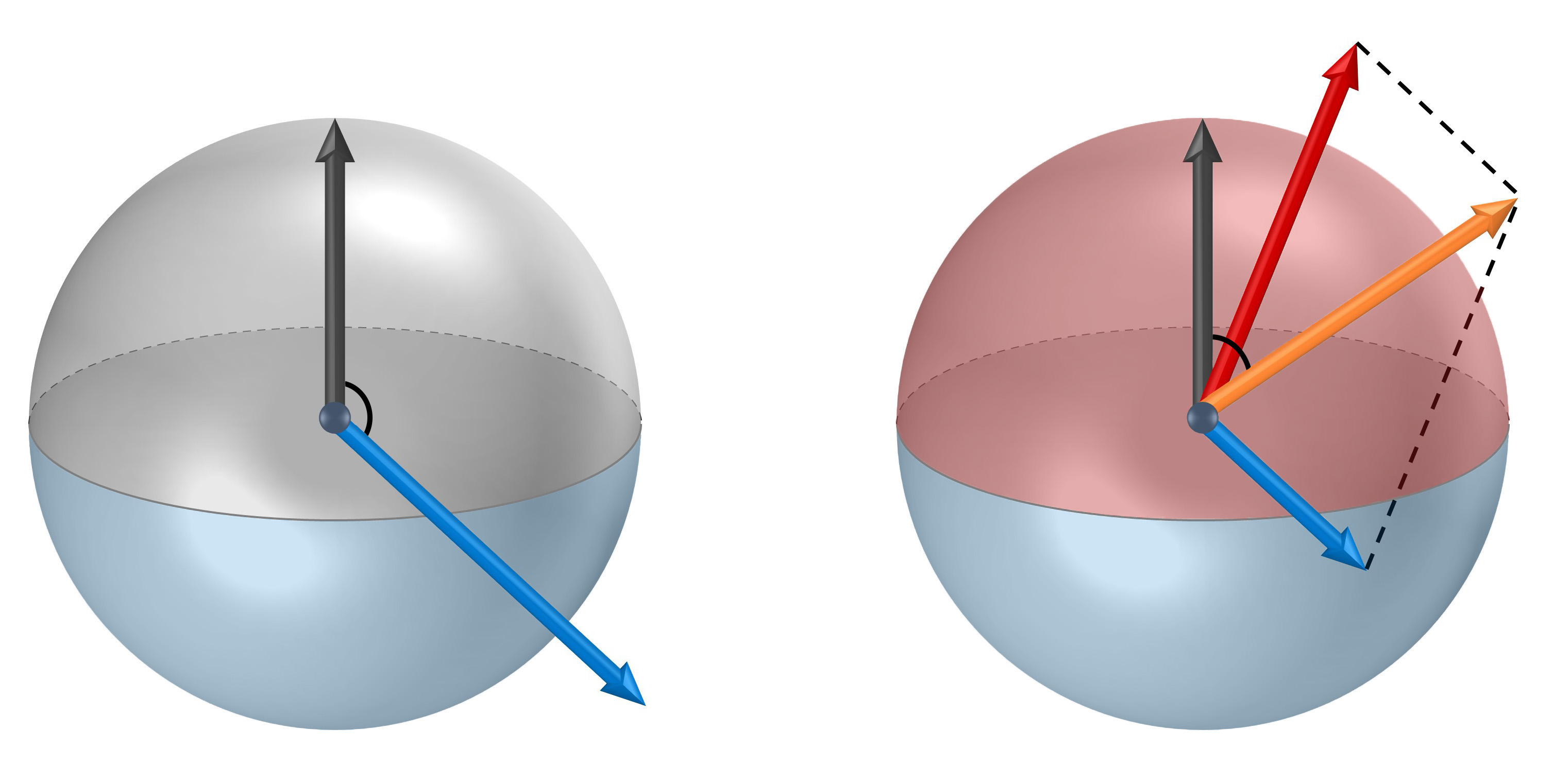}
        \put(0,44) {\fontsize{9}{1}\selectfont (\textit{a})}
        \put(55,44) {\fontsize{9}{1}\selectfont (\textit{b})}
        \put(20.5,42.5) {\fontsize{9}{1}\selectfont $\vec{s}$}
        \put(24,22.5) {\fontsize{9}{1}\selectfont $\theta\in[\frac{\pi}{2},\pi]$}
        \put(42,2) {\fontsize{9}{1}\selectfont $\vec{H}_{\mathrm{eff}}=\rmDelta\vec{s}$}
        \put(75.8,42.5) {\fontsize{9}{1}\selectfont $\vec{s}$}
        \put(88,11) {\fontsize{9}{1}\selectfont $\frac{1}{2}\rmDelta\vec{s}$}
        \put(86.2,46.5) {\fontsize{9}{1}\selectfont $\vec{m}'$}
        \put(90,37.5) {\fontsize{9}{1}\selectfont $\vec{H}_{\mathrm{eff}}=\frac{1}{2}\rmDelta\vec{s} + \vec{m}'$}
        \put(80,27) {\fontsize{9}{1}\selectfont $\theta\in[0,\pi]$}
    \end{overpic}
    \caption{Schematic of orientations of the spin vector $\vec{s}$ and the effective field $\vec{H}_{\mathrm{eff}}$ in (\textit{a}) the isotropic Heisenberg spin system and (\textit{b}) the spin Euler system.
    The variation of $\rmDelta\vec{s}$ is restricted to the southern hemisphere (shaded in blue), whereas $\vec{m}'$ is restricted to the northern hemisphere (shaded in red).}
    \label{fig:s_H}
\end{figure}

As sketched in figure~\ref{fig:s_H}, the range of the angle $\theta \equiv \arccos ((\vec{s} \cdot \vec{H}_{\mathrm{eff}})/|\vec{H}_{\mathrm{eff}}|)$ between $\vec{s}$ and $\vec{H}_{\mathrm{eff}}$ depends on the spin system.
The isotropic Heisenberg spin system has $\theta \in [\pi/2,\pi]$ with
\begin{equation}
    \vec{s} \cdot \vec{H}_{\mathrm{eff}}
    = \vec{s} \cdot \rmDelta \vec{s}
    = -|\bn \vec{s}|^2
    \le 0,
\end{equation}
whereas the spin Euler system has $\theta \in [0,\pi]$ with
\begin{equation}
    \vec{s} \cdot \vec{H}_{\mathrm{eff}}
    = \vec{s} \cdot \left(\frac{1}{2} \rmDelta \vec{s} + \vec{m}' \right)
    = |\vec{u}|^2 - \frac{1}{4} |\bn \vec{s}|^2.
\end{equation}
Presuming $\vec{H}_{\mathrm{eff}}$ is directed to the north pole,
$\vec{s}$ in the isotropic Heisenberg spin system is confined to the southern hemisphere, whereas there is no such restriction in the spin Euler system due to the additional term $\vec{m}'$. 
Therefore, the spin Euler system for ideal flows can have greater degrees of freedom and more complex dynamics than the isotropic Heisenberg spin system for magnetic crystals.

\section{Upper bound estimation for $\norm{\rmDelta\vec{s}}$}\label{app:Lap_s}

We show an attempt on estimating the upper bound of $|\rmDelta\vec{s}|$ which is an important ingredient in the non-blowup criterion in \eqref{eq:non-blowup_condition_LaplacianS}.
Taking the inner product of $\rmDelta\vec{s}$ and the Laplacian of \eqref{eq:spin_Euler} yields the evolution equation for $|\rmDelta\vec{s}|$
\begin{equation}\label{eq:time_rate_Laplacians}
    \p_t|\rmDelta\vec{s}|^2
    + 2\rmDelta\vec{s}\cdot\rmDelta(\vec{s}\times\vec{m})
    = 0.
\end{equation}

Integrating \eqref{eq:time_rate_Laplacians} over $\mathcal{D}$ yields
\begin{align}
    \p_t\norm{\rmDelta\vec{s}}_{2}^2
    =\ & -2\int_{\mathcal{D}} \varepsilon_{jrs}\rmDelta s_j\frac{\p}{\p x_k}\left(\frac{\p s_r}{\p x_k}m_s  + s_r\frac{\p m_s}{\p x_k}\right)\dif V
    \notag \\
    =\ & -2\oiint_{\p\mathcal{D}}\varepsilon_{jrs} n_k\rmDelta s_j \left(\frac{\p s_r}{\p x_k}m_s  + s_r\frac{\p m_s}{\p x_k}\right)\dif S
    \notag \\
    &+ 2\int_{\mathcal{D}}\varepsilon_{jrs}\frac{\p\rmDelta s_j}{\p x_k}\left(\frac{\p s_r}{\p x_k}m_s  + s_r\frac{\p m_s}{\p x_k}\right)\dif V
    \notag \\
    =\ & 2\oiint_{\p\mathcal{D}} \varepsilon_{jrs}n_k\frac{\p\rmDelta s_j}{\p x_k}s_rm_s\dif S
    - 2\int_{\mathcal{D}}\varepsilon_{jrs}\rmDelta^2 s_js_rm_s\dif V
    \notag \\\
    =\ & -2\int_{\mathcal{D}} \rmDelta^2\vec{s}\cdot(\vec{s}\times\vec{m})\dif V.
    \label{eq:ineq_LapsL2_Lap2s}
\end{align}
Applying the H\"older inequality to \eqref{eq:ineq_LapsL2_Lap2s} we obtain
\begin{align}
    \p_t\norm{\rmDelta\vec{s}}_{2}^2
    &\le \frac{1}{2}\left( \sqrt{6}\norm{\rmDelta\vec{s}}_{1} + \sqrt{2}C_\omega\norm{\rmDelta\vec{s}}_{2}^2\right)\norm{\rmDelta^2\vec{s}}_{\infty}
    \notag \\
    &\le \frac{1}{2}\left(\sqrt{6}\mu^{1/2}(\mathcal{D})\norm{\rmDelta\vec{s}}_{2} +  \sqrt{2}C_\omega\norm{\rmDelta\vec{s}}_{2}^2\right)\norm{\rmDelta^2\vec{s}}_{\infty},
\end{align}
which yields
\begin{equation}\label{eq:ineq_LapsL2_Lap2s_2}
    \p_t\norm{\rmDelta\vec{s}}_{2}
    \le \frac{1}{4}\left(\sqrt{6}\mu^{1/2}(\mathcal{D}) + \sqrt{2}C_\omega\norm{\rmDelta\vec{s}}_{2} \right)\norm{\rmDelta^2\vec{s}}_{\infty}.
\end{equation}
Integrating \eqref{eq:ineq_LapsL2_Lap2s_2} over time yields
\begin{equation}\label{eq:ineq_LapsL2_Lap2s_3}
    \norm{\rmDelta\vec{s}}_{2}
    \le C_{s1}\exp\left(\frac{\sqrt{2}C_\omega}{4}\int_0^t \norm{\rmDelta^2\vec{s}(\cdot,\tau)}_{\infty}\dif\tau \right) + C_{s2},
\end{equation}
with constants
\begin{equation}
    C_{s1} = \frac{1}{\sqrt{2}C_\omega}\left(\sqrt{6}\mu^{1/2}(\mathcal{D}) + \sqrt{2}C_\omega\norm{\rmDelta\vec{s}(\cdot,0)}_{2}\right), \quad
    C_{s2} = -\frac{\sqrt{3}\mu^{1/2}(\mathcal{D})}{C_\omega}.
\end{equation}

The inequality \eqref{eq:ineq_LapsL2_Lap2s_3} implies a closure problem in the bound estimation -- the growth of the $L^2$-norm of $\rmDelta\vec{s}$ depends on its higher order derivatives.
In addition, using the identity
\begin{equation}
    \rmDelta\vec{s}\cdot\rmDelta(\vec{s}\times\vec{m})
    = \rmDelta\vec{s}\cdot(\vec{s}\times\rmDelta\vec{m}) + 2\bn\vec{s}:(\bn\vec{m}\times\rmDelta\vec{s})
\end{equation}
we estimate the growth rate of the $L^\infty$-norm of $\rmDelta\vec{s}$
\begin{equation}\label{eq:ineq_LapsLinfty}
    \p_t\norm{\rmDelta\vec{s}}_\infty
    \le \norm{\rmDelta\vec{m}}_\infty + \norm{\bn\vec{m}}_\infty\norm{\rmDelta\vec{s}}_\infty^{1/2}.
\end{equation}
However, it appears to be challenging to estimate $\norm{\rmDelta\vec{m}}_\infty$ and $\norm{\bn\vec{m}}_\infty$ in terms of $\norm{\rmDelta\vec{s}}_\infty$.
Thus, the estimations of \eqref{eq:ineq_LapsL2_Lap2s_3} and \eqref{eq:ineq_LapsLinfty} need to be improved in the future work.

\bibliographystyle{jfm}

\end{document}